\documentclass[pra,twocolumn,showpacs,superscriptaddress]{revtex4-1}

\usepackage{epsf,epsfig}
\usepackage[psamsfonts]{amssymb}
\usepackage{amsmath}
\usepackage{bm}
\usepackage{natbib}
\usepackage{graphicx}
\usepackage{color}
\usepackage{mathrsfs}
\def\imath{\rm i}
\def\G{\mathord{\buildrel{\lower3pt\hbox{$\scriptscriptstyle\leftrightarrow$}}
\over G}}

\newcommand{\bfsfG}{\mbox{\sffamily\bfseries{G}}}
\newcommand{\sssection}[1]{{\par\it #1.---}}

\begin{document}
\title{Quantum optical effective-medium theory for layered metamaterials}
\author{Ehsan Amooghorban}
\email{ehsan.amooghorban@sci.sku.ac.ir}
\affiliation{Department of Physics, Faculty of Basic Sciences, Shahrekord University, P.O. Box 115, Shahrekord 88186-34141, Iran}
\affiliation{Photonics Research Group, Shahrekord University, P.O. Box 115, Shahrekord 88186-34141, Iran}

\author{Martijn Wubs}
\email{mwubs@fotonik.dtu.dk}
\affiliation{Department of Photonics Engineering, Technical University of Denmark, DK-2800 Kgs. Lyngby, Denmark}
\affiliation{Center for Nanostructured Graphene, Technical University of Denmark, DK-2800 Kgs. Lyngby, Denmark}
% %%%%%%%%%%%%%%%%%%%%%%%%%%%%%%%%%%%%%%%%%%%%%%%%%%%%%%%%%%%%%%%%%%
\begin{abstract}
The quantum optics of metamaterials starts with the question whether the same effective-medium theories apply as in classical optics.  In general the answer is negative. For active plasmonics but also for some passive metamaterials, we show that an additional effective-medium parameter is indispensable besides the effective index, namely the effective noise-photon distribution. Only with the  extra parameter  can one predict how well the quantumness of states of light is preserved in the metamaterial. The fact that the effective index alone is not always sufficient and that one additional effective parameter suffices in the quantum optics of metamaterials is both of fundamental and practical interest.  Here from a Lagrangian description of the quantum electrodynamics of media with both linear gain and loss, we compute the effective noise-photon distribution for quantum light propagation in arbitrary directions in layered metamaterials, thereby detailing and generalizing our recent work [ E. Amooghorban {\em et al.}, Phys. Rev. Lett. {\bf 110}, 153602 (2013)]. The effective index with its direction and polarization dependence is the same as in classical effective-medium theories. As our main result we derive both for passive and for active media how the value of the effective noise-photon distribution  too depends on the polarization and propagation directions of the light. Interestingly, for TE-polarized light incident on passive metamaterials, the noise-photon distribution reduces to a thermal distribution, but for TM-polarized light it does not.  We illustrate the robustness of our quantum optical effective-medium theory by accurate predictions both for power spectra and for balanced homodyne detection of output quantum states of the metamaterial.
\end{abstract}


%\keywords{}

\date{\today}

% %%%%%%%%%%%%%%%%%%%%%%%%%%%%%%%%%%%%%%%%%%%%%%%%%%%%%%%%%%%%%%%%%%

\maketitle

% %%%%%%%%%%%%%%%%%%%%%%%%%%%%%%%%%%%%%%%%%%%%%%%%%%%%%%%%%%%%%%%%%%

\section{Introduction}\label{Sec:Introduction}

% Introduction

% metamaterials and metasurfaces
Metamaterials are known and studied for guiding and manipulating light in ways not seen in Nature.~\cite{Pendry2000a,Shalaev2007a}. They consist of repeated designed subwavelength unit-cell structures that allow a description of the metamaterial in terms of effective optical parameters not found in natural materials, with negative-index metamaterials~\cite{Pendry2000a,Shalaev:2005a} as the prime example.
In this Introduction we discuss applications of metamaterials in quantum optics, and motivate the need for a quantum optical effective-medium theory.

Applications in optics of metamaterials include flat superlenses~\cite{Pendry2000a,Ramakrishna:2003a,Wood2006a,Yan:2013a} and sensors~\cite{Chen:2012a}.  Metamaterials can constitute a material basis for applications of transformation optics~\cite{Pendry:2015a} such as cloaking devices, which typically require graded-index  media realized as graded-{\em effective}-index media.
The properties of a metamaterial derive from an average of its constituting materials, which often involve both metals and dielectrics. There are different ways to determine the effective refractive index of a metamaterial, which is the topic of homogenization theory~\cite{Bergman:1978a,Smith2002a,Smith2006a,Acher2007a,Sun2009a,Felbacq2009a,Menzel2008a,Andryieuski2009a,Mortensen2010a,Papadakis:2015a}.

One important class of structures for which such averaging can produce  truly new functionalities are the epsilon-near-zero (or ENZ) materials~\cite{Ziolkowski2004a,Silveirinha2006a,Edwards:2008a,Sokhoyan:2013a}, in which light propagates with extremely small phases and long effective wavelengths, as has been realized also at visible wavelengths~\cite{Moitra2013a,Maas2013a}. Dispersion-compensated metamaterials can also lead to new devices~\cite{Jiang:2014a}.
Loss-compensated metamaterials constitute another class of structures for which averaging over a unit cell can produce something truly new~\cite{Ramakrishna:2003a,Grgic:2012a,Manzanares:2014a}: loss in one constituent can be compensated  by linear gain in another, so as to produce metamaterials with lower or even vanishing effective loss.
Partial loss compensation has been realized both in plasmonic waveguides~\cite{DeLeon2009a,Berini2012a} and in metamaterials~\cite{Xiao2010a}.
Loss compensation is studied in the field of active plasmonics and tuneable metamaterials~\cite{Boardman2011a,Zayats2013a,Jung:2014a}. All mentioned applications of metamaterials are within the realm of classical electromagnetism.

% ??
Quantum plasmonics concerns the study of quantum optics with plasmons~\cite{Tame2013a}. It is a stimulating question which of the mentioned applications of metamaterials can be transferred to quantum optics.  Indeed an increasing number of researchers is exploring how to manipulate quantum emitters and quantum states of light using metamaterials~\cite{Cortes:2012a,Siomau:2012a,Wang:2012a,Lu:2014a,Jha:2015a,Roger:2015a,Abdullah:2015a,Asano:2015a,Zhang:2016a}.   Vice versa, the exploration how quantum states of light can be used to analyze metamaterial properties has also only just begun~\cite{Zhou:2012a,Amooghorban2013a}.

The best known and important example of metamaterials with new functionality for quantum emitters are the hyperbolic metamaterials. Their effective epsilon is positive in one or two directions and negative-valued in the remaining direction(s)~\cite{Cortes:2012a}. By taking the usual limit of infinitely small unit cells, the iso-frequency dispersion surfaces of such anisotropic bulk media become hyperbolic, with infinite associated local optical density of states. This nonphysical infinity indicates that the usual idealized description of metamaterials needs improvement for embedded quantum emitters, for example by taking into account the nonlocality of the metallic response~\cite{Yan2012a}, or the finite size of either the unit cells~\cite{Jacob2010} or the emitters~\cite{Poddubny:2011a}.  Thus quantum emitters embedded inside metamaterials provide a challenge for the effective-medium theories~\cite{Kidway:2012a}.

Quantum optics poses another less known challenge to metamaterials, even when probing metamaterials in the far field and when unit cells are much smaller than the operating wavelength:
One can  do quantum optical experiments to tell apart two metamaterials even though they have the same shape and the same effective index~\cite{Amooghorban2013a}. In classical electrodynamics this would be impossible, but in quantum optics this may even be possible with normally incident light on simple layered metamaterials~\cite{Amooghorban2013a}. This is because of quantum noise. Quantum mechanics poses a limit to the use of the common  effective-index theories.

In principle the `quantumness' of light can survive the propagation through a metamaterial. In general, quantum states of light that propagate through absorbing or amplifying media will be affected by quantum noise associated with the loss~\cite{Huttner1992a,Huttner1992b,Gruner1996a,Wubs2001a,Suttorp2004a} and gain~\cite{Matloob1997a,Scheel1998a}. This also applies to metamaterials. This  does not mean that the concept of the effective index breaks down in quantum optics.
On the contrary, in Ref.~\onlinecite{Amooghorban2013a} we presented a quantum optical effective-index theory that accurately describes passive metamaterials and the more exotic metamaterials consisting of alternating layers both with gain. The theory also describes the quantum noise in these metamaterials, and can be seen as a direct extension of the usual effective-index theory.

However, for loss-compensated metamaterials we found that the effective index sometimes underestimates the average quantum noise picked up in a unit cell, because loss can be compensated by gain but quantum noise due to loss cannot be compensated by quantum noise due to gain. Thus effective descriptions of loss-compensated metamaterials based solely on the effective index break down in quantum optics.
Nevertheless an accurate quantum optical effective-medium theory  of loss-compensated metamaterials is still possible, where besides the usual effective index an additional effective-medium parameter is introduced for loss-compensated metamaterials, namely the effective noise photon distribution~\cite{Amooghorban2013a}. These results were obtained only for normally incident light on multilayer metamaterials.

Here we generalize Ref.~\cite{Amooghorban2013a} in important ways by considering quantum optical effective-medium theories for {\em three-dimensional} light propagation in layered metamaterials. As is well known in classical optics, TE- and TM-polarized light propagate qualitatively different in a layered medium. Analogously, we will here present surprisingly different effective noise-photon densities for TE- and TM-polarized light. Only for normal incidence will they coincide with each other and with the effective noise-photon density of the one-dimensional theory of Ref.~\cite{Amooghorban2013a}. We also address anew the question whether it is only the loss-compensated metamaterials that require an additional effective-medium parameter.

% organization of paper
 The paper is organized as follows: In Sec.~\ref{Sec:quantization}, we introduce the field quantization of media with both gain and loss, presenting what we believe is the shortest and simplest route from a Lagrangian to a phenomenological quantum electrodynamics based on the classical Green function. We use these results to derive in Sec.~\ref{Sec:Multilayer} an input-output relation for planar multilayer dielectrics.
 Then in Sec.~\ref{Sec:effective_index} we derive a quantum optical effective-index theory, and we test its predicted power spectra in Sec.~\ref{sec:intensity}. A quantum optical effective-medium theory for both $s$- and $p$-polarized light is introduced in Sec.~\ref{Sec:effective_medium}, and tested for the propagation of squeezed states of light through  metamaterials in Sec.~\ref{Sec:squeeze}. We end with a discussion and conclusions  in Sec.~\ref{Sec:Conclusions}, and an appendix.

%***************************************

\section{Field quantization}\label{Sec:quantization}
With application to loss-compensated metamaterials in mind, here we derive a general expression for the quantized electric field  after non-normal propagation through a bounded inhomogeneous dielectric medium that exhibits both loss and gain. Quantum-mechanical theories for electromagnetic wave propagation
through lossy~\cite{Huttner1992a,Huttner1992b,Gruner1996a} or amplifying~\cite{Matloob1997a,Scheel1998a} dielectrics have been developed previously. We described media with both gain and loss in Ref.~\onlinecite{Amooghorban2011a}, where we used path-integral quantization techniques. Here instead we will not use path integrals and instead we give a simpler quantum electrodynamical description of media with both gain and loss, which is valid for arbitrary dielectric structures, including all non-magnetic metamaterials. The method  has the advantage that there is a clear relation between the dielectric function of the dielectric medium and the more microscopic coupling parameters in the Lagrangian.  Its specific application to multilayer structures then follows in Sec.~\ref{Sec:Multilayer}.

The quantum electrodynamics of a linearly lossy dielectric
can be described by modeling the medium as a reservoir of three-dimensional harmonic
oscillators that interacts with the electromagnetic field~\cite{Huttner1992a}. We also allow for the possibility that the medium is linearly amplifying in some finite regions of space, with gain ($\mbox{Im}[\varepsilon(\omega)]\equiv \varepsilon_{\rm I}(\omega)<0$) in one or more finite-frequency windows. Linear gain can be modeled as the coupling of the electromagnetic field to a continuum of inverted harmonic oscillators~\cite{Glauber1986a,Gardiner2004a}.

We introduce our model for optical media with both gain and loss by first specifying its Lagrangian density in real space~\cite{Amooghorban2011a}
\begin{equation}\label{total Lagrangian}
{\cal L}={\cal L}_{\rm EM}+{\cal L}_{\rm e}+{\cal L}_{\rm int},
\end{equation}
where the first term ${\cal L}_{\rm EM}$ has the standard
form
$
{\cal L}_{\rm EM}=\frac{1}{2}\varepsilon_0{\bf E}^2({\bf x},t)-\frac{1}{2\mu_0}{\bf
B}^2({\bf x},t),
$
describing the free electromagnetic field. There is gauge freedom to write the electric field ${\bf E} = -
\partial {\bf A}/\partial t - \nabla \phi$ and the
magnetic field ${\bf B} = \nabla \times {\bf A} $ in terms of the
scalar and vector potentials $\phi$ and $\bf A$.  For convenience we
choose the Coulomb gauge in which the divergence of the vector potential vanishes by definition.
%, which
%allows us to write ${\bf E}$ and ${\bf B}$ in terms of only the
%vector potential.
%
The second term ${\cal L}_{\rm e}$ in Eq.~(\ref{total Lagrangian}) denotes the internal dynamics of the linear medium, which we describe in terms of frequency continua of the harmonic vector field ${\bf X}_\omega({\bf x},t)$ as
\begin{eqnarray}\label{medium Lagrangian}
{\cal L}_{\rm e}&=&\frac{1}{2}\int_0^\infty \mbox{d}\omega
\,\left[{{\dot{\bf X}^2}_\omega({\bf
x},t)-\omega^2{\bf X}^2_\omega({\bf
x},t)}\right]\,{\rm sgn}[ \varepsilon_{\rm I}({\bf x},\omega)].\,\,\,\,\,\,\,
\end{eqnarray}
We define the polarization field of the medium as
\begin{eqnarray}\label{definition of polarization}
{\bf P}({\bf x},t)=\int_0^\infty \mbox{d}\omega \, g({\bf x},\omega){\bf
X}_\omega({\bf x},t),
\end{eqnarray}
and assume a linear coupling of the electromagnetic field with this field,
\begin{equation}\label{interaction Lagrangian}
{\cal L}_{\rm int}({\bf A},{\bf P},\phi) = {\bf A}({\bf
x},t)\cdot \dot{\bf P}({\bf x},t)+\phi \nabla\cdot {\bf P}.
\end{equation}
The $g({\bf x},\omega) $  in Eq.~(\ref{definition of polarization}) is assumed to be a real-valued scalar
coupling function of the electromagnetic field to the spatially inhomogeneous medium.
At positions and for frequencies for which $\varepsilon_{\rm I}({\bf x},\omega)$ is positive-valued, the medium is lossy and ${\bf X}^{2}_{\omega}({\bf x},t)$ is an oscillator to which electromagnetic energy is lost, whereas if $\varepsilon_{\rm I}({\bf x},\omega)$ has a negative value, then the medium is amplifying the electromagnetic signal. The latter is modeled with oscillators that are called `inverted' because of the overall minus sign ${\rm sgn}[ \varepsilon_{\rm I}({\bf x},\omega)] = -1$ in the material Lagrangian density Eq.~(\ref{medium Lagrangian}).
 The time derivative of the scalar potential
($\dot{\phi}$) does not appear in the Lagrangian density~(\ref{total Lagrangian}). This implies in the first place that the conjugate momentum associated with the scalar potential $\phi$ is identically zero. Secondly, the scalar potential (by its Euler-Lagrange equation) can be expressed in terms of other  degrees of freedom by Poisson's equation $\varepsilon_0\nabla^2 \phi= \nabla\cdot {\bf P} $. The solution is
$
\phi({\bf x},t)=(4\pi \varepsilon_0)^{-1}\int \mbox{d}{\bf x}' \nabla'\cdot {\bf P}({\bf x}',t)/|{\bf x}-{\bf x}'|
$.
The scalar potential is thereby eliminated, and  a reduced Lagrangian is obtained where only the vector potential {\bf A}, the harmonic vector field ${\bf X}_\omega$ and their time derivatives appear. To this end, the free electromagnetic field part and its interaction part are rewritten as
\begin{subequations}
\begin{eqnarray}
{\cal L}_{\rm EM}({\bf A}) & =& \frac{1}{2}\varepsilon_0\dot{{\bf A}}^2({\bf x},t)-\frac{1}{2\mu_0}({\nabla \times {\bf
A}}({\bf x},t))^2, \label{New free electromagnetic Lagrangian} \\
{\cal L}_{\rm int}({\bf A},{\bf P}) &  = & {\bf A}({\bf
x},t)\cdot \dot{\bf P}({\bf x},t) \nonumber \\
& +& \frac{1}{8\pi \varepsilon_0}\int \mbox{d}{\bf x}' \frac{\nabla\cdot {\bf P} ({\bf x},t) \nabla'\cdot {\bf P}({\bf x}',t)}{|{\bf x}-{\bf x}'|}, \label{New interaction Lagrangian}
\end{eqnarray}
\end{subequations}
while the material Lagrangian density~(\ref{medium Lagrangian}) stays without any changes because there is no term including the scalar potential $\phi$. Here and in the following we take the medium to be non-magnetic, and for extensions to magnetodielectrics we refer to Ref.~\onlinecite{Amooghorban2011a}.
The Lagrangian~(\ref{total Lagrangian}),
with the vector potential ${\bf A}$, and the continua of the
polarization operator ${\bf X}_\omega$ can be used as
canonical fields with the following corresponding canonically conjugate fields
\begin{subequations}\label{canonically conjugate fields}
\begin{eqnarray}
- \varepsilon_0 {\bf E}({\bf x},t)& \equiv& \frac{\delta {\cal L}}{\delta \dot{\bf
{A}}({\bf x},t)}= \varepsilon_0{\dot {\bf A}}({\bf x},t), \label{canonically conjugate fields_a}\\
{\bf Q}_\omega({\bf x},t) &\equiv& \frac{\delta {\cal L}}{\delta \dot{\bf
{X}_\omega}({\bf x},t)}  \\
& = & g(\omega,{\bf x}){\bf A}({\bf x},t)+ {\rm sgn}[\varepsilon_{\rm I}({\bf x},\omega)]{\dot
{\bf X}}_\omega({\bf x},t). \nonumber  \label{canonically conjugate fields_b}
\end{eqnarray}
\end{subequations}
Until now there is no difference with a classical description. We arrive at a quantum theory by taking the  fields to be quantum fields (operator vector fields) that satisfy non-vanishing equal-time commutation relations with their canonically conjugate fields. Apart from the subtlety with the sign functions in Eq.~(\ref{canonically conjugate fields_b}) that discriminate
between the frequency intervals where there is gain and loss, this
canonical quantization of the  fields can proceed in a standard
fashion by demanding equal-time commutation relations
\begin{subequations}\label{equal_time_commutation}
\begin{eqnarray}
\left[ { {A} _i ({\bf x},t),-\varepsilon_0 {E}_j ({\bf x'},t)} \right] &=& {\rm
i} \hbar \,\delta _{ij} \delta^{\bot} ({\bf x} - {\bf x'}),\label{b2}\\
\left[ { {X} _{\omega,i} ({\bf x},t), {Q}_{\omega',j} ({\bf x'},t)}
\right]& =& {\rm i}\hbar\, \,\delta
_{ij}\,\delta(\omega-\omega')\, \delta^{3} ({\bf x} - {\bf x'}),\;\;\,\,\,\,\,\,\,\,\,
\end{eqnarray}
\end{subequations}
while all other equal-time commutators vanish. Using the Lagrangian~(\ref{total Lagrangian})
and the expression for the canonical conjugate variables in
Eq.~(\ref{canonically conjugate fields}), we obtain the Hamiltonian density
\begin{eqnarray} \label{Hamiltonian density}
{\cal H}({\bf x},t) &=&  \frac{1}{2}\varepsilon_0{\bf E}^2({\bf
x},t)+\frac{{\bf B}^2({\bf x},t)}{2\mu_0}\\
&&+\frac{1}{2}\int_0^\infty \mbox{d}\omega\, {\rm
sgn}[\varepsilon_{\rm I}({\bf x},\omega)]
\nonumber \\
&&\times \bigl\{ ({\bf Q}_\omega({\bf x},t)-g(\omega,{\bf x}){\bf A}({\bf x},t))^2+\omega ^2 {\bf X}_\omega^2({\bf
x},t) \bigl\}.\nonumber
\end{eqnarray}
Maxwell's equations can now be obtained from the Heisenberg
equations of motion for the vector potential and the transverse electric field and from the commutation relation Eq.~(\ref{equal_time_commutation}),
\begin{subequations}\label{ADderivatives}
\begin{eqnarray}
\dot{\bf {A}}({\bf x},t) & = & - {\bf E}({\bf x},t), \\
\varepsilon_0\dot{\bf {E}}({\bf x},t) & = & \mu_{0}^{-1}\nabla\times\nabla\times{\bf
A}({\bf x},t) -\dot{{\bf P}}({\bf x},t).
\end{eqnarray}
\end{subequations}
Using the definitions ${\bf D} =  \varepsilon_0{{\bf E}}+{{\bf P}}$  for the displacement field and and  ${\bf H} = {\bf B}/\mu_0$ for the magnetic field strength,  Eqs.~(\ref{ADderivatives}) result in $\dot{\bf {D}}({\bf
x},t)=\nabla\times{\bf H}({\bf x},t)$ and $\dot{\bf {B}}({\bf
x},t)=-\nabla\times{\bf E}({\bf x},t)$, showing the consistency with Maxwell's equations. In a similar
fashion, the Heisenberg equation of motion for the dynamical
variable ${\bf X}_{\omega}$ leads to the second-order differential equation
\begin{eqnarray}
\ddot{\bf {X}}_\omega({\bf x},t) &  = & -\omega^2{\bf X}_\omega({\bf
x},t) + {\rm sgn}[\varepsilon_{\rm
I}(\omega)] g({\bf x},\omega){\bf E}({\bf x},t),\hspace{.8cm}
\end{eqnarray}
which has the formal solution
\begin{eqnarray}\label{X_formally_solved}
{\bf X}_\omega ({\bf x},t)  &= & \left({\dot{\bf  X}_\omega ({\bf
x},0)\frac{{\sin \omega t}}{\omega }
+ {\bf X} _\omega ({\bf
x},0)\cos \omega t}\right)  \\
& + & g({\bf x},\omega){\rm
sgn}[\varepsilon_{\rm I}({\bf x},\omega)]\int_0^t \mbox{d}t' \frac{{\sin \omega
(t - t')}}{\omega } {\bf {\bf E}}({\bf
x},t').\nonumber
\end{eqnarray}
In classical electrodynamics, one would typically assume the corresponding initial fields
${\dot{\bf  X}}_{\omega}({\bf x},0)$ and ${\bf X}_{\omega}({\bf x},0)$ to vanish, which is something that one should not do for the initial quantum operators in Eq.~(\ref{X_formally_solved}) if only because this would violate their commutation relations. It is these initial-operator terms in Eq.~(\ref{X_formally_solved}) that give rise to the qualitatively new phenomenon of quantum noise, as we shall see shortly.

To facilitate our further calculations, let us introduce the annihilation operator
\begin{equation}\label{creationandannihilation}
d_j ({\bf x},\omega ,t) = \frac{1}{ \sqrt {{2\hbar\omega }}}
\left[ -{\rm i} \omega {\rm X}_{\omega,j}({\bf x},t ) + \,{\rm Q}_{\omega,j}({\bf x},t) \right],
\end{equation}
where $j=1,2,3$ labels the three orthogonal spatial directions. Their commutation relations follow immediately from Eq.~(\ref{equal_time_commutation}),
\begin{eqnarray}\label{comm_creation_annihilation}
\left[ {d_j  ({\bf x},\omega ,t),d_{j'}^ \dag ({\bf x}',\omega ',t)}
\right] & = & \,\delta _{jj'}\,\delta (\omega  - \omega ')\delta^{3}({\bf
x} - {\bf x}').\,\,\,\,\,\,\,
\end{eqnarray}
Now by inverting the relations~(\ref{creationandannihilation}) and substituting the result into Eq.~(\ref{definition of polarization}), the polarization field of the medium can be written in terms of creation and annihilation operators as
\begin{eqnarray}\label{PandM}
{\bf P}({\bf x},t)&=&\varepsilon_0\int_0^\infty
\mbox{d}t'\,\chi({\bf x},t-t'){\bf E}({\bf x},t')+{\bf
P}^{\rm N}({\bf x},t).\,\,\,\,\,
\end{eqnarray}
Here, the time-dependent susceptibility is defined as
\begin{equation}\label{Susceptibility definition}
\chi({\bf x},t)=\frac{\Theta (t)}{\varepsilon_0}\int_0^\infty \mbox{d}\omega\, {\rm sgn}[\varepsilon_{\rm I}({\bf x},\omega)]g^2({\bf x},\omega)\frac{{\sin \omega t}}{\omega },
\end{equation}
which is a causal response function because of the step function $\Theta (t)$. After Fourier transformation, the susceptibility becomes
\begin{equation}
\chi({\bf x},\omega)
=\frac{1}{\varepsilon_0}\int_0^\infty \mbox{d}\omega'
\frac{g^{2}({\bf x},\omega' )\,{\rm sgn}[\varepsilon_{\rm
I}({\bf x},\omega')]}{\omega'^2-(\omega+\imath0^+)^2}.
\end{equation}
The field ${\bf P}^{\rm N}({\bf x},t)$ in Eq.~(\ref{PandM}) is the electric polarization noise density that is inevitably associated with absorption and amplification inside medium. As in
the phenomenological method of Refs.~\onlinecite{Matloob1997a,Scheel1998a}, we can
separate this noise operator into positive- and negative-frequency
parts ${\bf P}^{\rm N}={\bf P}^{\rm N (+)}+{\bf P}^{\rm N(-)}$ with
${\bf P}^{\rm N(-)} = [{\bf P}^{\rm N(+)}]^{\dag}$, where
\begin{eqnarray}\label{Noiseoperator}
{P}^{\rm N (+)}_i({\bf x},t)= {\rm i}\int_0^\infty \mbox{d}\omega\,
\sqrt{\frac{\hbar\varepsilon_0|\varepsilon_{\rm
I}({\bf x},\omega)}|{\pi}}f_i({\bf x},\omega)e^{-{\rm i}
\omega t},\,\,\,\,\,\,\,\,\,
\end{eqnarray}
in terms of the operator $ f_i({\bf x},\omega)$ that
has the form ${d}_i({\bf x},\omega,0)\Theta[\varepsilon_{\rm I}({\bf x},\omega)]+{d}_i^\dag({\bf
x},\omega,0)\Theta[-\varepsilon_{\rm I}({\bf x},\omega)]$. This noise operator is indeed expressed in terms of material operators at the initial time $t=0$, as anticipated.
If we now take the time derivative of Maxwell's equations in Eq.~(\ref{ADderivatives}) and insert Eq.~(\ref{PandM}), then we obtain the frequency-domain wave equation for the positive-frequency part of the vector potential
\begin{eqnarray}\label{wave_equation_with_noise}
&&\nabla \times  \nabla  \times {\bf
A}^{ (+)}
%({\bf x},\omega)
- \frac{\omega ^2}{c^2}
\varepsilon
%({\bf x},\omega)
{\bf A}^{ (+)}
%({\bf x},\omega)
=-{\rm i} \mu_0\omega{\bf P}^{\rm N (+)}
%({\bf x},\omega)
,
\end{eqnarray}
where  the electric permittivity $\varepsilon({\bf x},\omega) = 1 + \chi({\bf x},\omega)$ satisfies the Kramers-Kronig relations because $\chi({\bf x},t)$ in Eq.~(\ref{Susceptibility definition}) vanishes for $t< 0$. Furthermore, the noise operator ${\bf P}^{\rm N (+)}({\bf x},\omega)$ in the wave equation~(\ref{wave_equation_with_noise}) plays the role of a Langevin force associated with the quantum noise sources in the dielectric. This equation can be solved as
\begin{eqnarray}\label{vector potential field}
{\bf A}^{ (+)} ({\bf x},t)&=&\frac{-{\rm i} \mu_0}{\sqrt{2\pi}}\int_0^\infty \mbox{d}\omega\,\omega
\int \mbox{d}^3 {\bf x'}\nonumber\\
&&\times\bfsfG({\bf x},{\bf x'},\omega) \cdot
{\bf P}^{\rm N (+)}({\bf x',\omega})\, e^{-{\rm i}\omega t}\nonumber\\
&=& \int_0^\infty \mbox{d}\omega \int \mbox{d}^3 {\bf x'}\, \sqrt{\frac{\hbar\mu_0 \omega^2 |{\varepsilon_{\rm
I}({\bf x'},\omega)}|}{\pi c^2}} \nonumber\\
&&\times\bfsfG({\bf x},{\bf x'},\omega) \cdot
{\bf f} ({\bf x'},\omega) e^{-{\rm i}\omega t},
\end{eqnarray}
where $\bfsfG({\bf x},{\bf x'},\omega)$ is the classical causal Green function (a tensor) that is defined by the equation
\begin{eqnarray}\label{Green equation}
&&\left[{\nabla \times  \nabla  \times - \frac{\omega ^2}{c^2}
\varepsilon({\bf x},\omega)} \right]\bfsfG({\bf x},{\bf x'},\omega) =\delta^{3}({\bf x}-{\bf x}'){\bm 1}_{3}.\,\,\,\,\,\,
\end{eqnarray}
From our Lagrangian theory we thus arrive at the following more phenomenological quantum theory of light in a medium with loss and gain: given a dielectric function $\varepsilon({\bf x},\omega)$, compute the classical Green function~(\ref{Green equation}) and use this to determine the vector potential~(\ref{vector potential field}). With Maxwell's equations all other fields can then also be determined.

\section{Input-output relation for planar dielectrics}\label{Sec:Multilayer}
\begin{figure}[t]
\includegraphics[width=1.0\columnwidth]{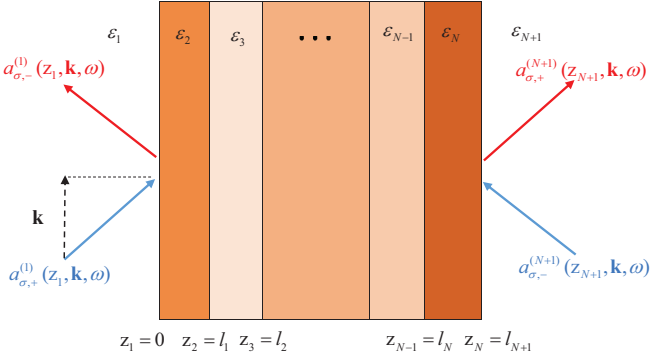}
\caption{(Color online)
Sketch of the planar dielectric medium with permittivity
$\varepsilon_j(\omega)$ and thickness $l_j$ of the $j^{\rm th}$ layer.  The arrows denote incoming and outgoing
fields. Also shown are the corresponding annihilation operators used
in the definitions of the electric-field operator Eq.~(\ref{Electric field in term of bosonic operator}).
}
\label{Fig:Scheme of the multilayer dielectric}
\end{figure}

Let us now specify that the dielectric medium with loss and/or gain is a planar dielectric for which the dielectric function $\varepsilon({\bf x},\omega)$ varies in a step-wise fashion in the $z$-direction, as depicted in Fig.~\ref{Fig:Scheme of the multilayer dielectric}.
Main goal of this paper is proposing and testing effective-medium theories that are accurate in quantum optics. The test consists of a comparison between an exact formalism for quantum optics in multilayer media on the one hand and  effective descriptions for  planar metamaterials on the other. In this section we will derive the mentioned exact formalism, while the numerical comparison will be made in  Sec.~\ref{Sec:effective_index}.

We look for a quantum optical input-output relation that can describe the action of a lossy and / or linearly amplifying multilayer medium on an arbitrary quantum state of light incoming from an arbitrary direction with either $s$- or $p$-polarization. The sought input-output relation is an operator relation, relating the (annihilation) operators describing the incoming light to the annihilation operators of the outgoing light. To this end, more is needed than just the classical scattering matrix of a multilayer medium,  because the incoming quantum state of light is not the only source that determines the output radiation. Quantum noise photons in the lossy and amplifying layers constitutes another source.

The classical Green function plays a central role, describing both the propagation of the incoming light and of the quantum noise photons towards the output direction or detector. As a generalization of the result by Toma{\v s} for lossy dielectric multilayers~\cite{Tomas1995a}, we already derived the Green function for multilayer media with both lossy and amplifying layers in Ref.~\onlinecite{Amooghorban2011a}, and we will make use of that result here. We will then show in more detail  that the explicit form of that Green function allows a suitable rewriting of the vector potential in Eq.~(\ref{vector potential field}) as the sought operator input-output relation. We will thereby arrive at a gain-and-loss-in-3D generalization of the 1D-formalism by Gruner {\em et al.} who studied the QED only of lossy planar dielectrics~\cite{Gruner1996a}. In our notation we follow Ref.~\cite{Tomas1995a}.

Multilayer media (see illustration in Fig.~\ref{Fig:Scheme of the multilayer dielectric}) are translationally invariant in two spatial directions.  It is often convenient to  exploit this symmetry by introducing the transverse spatial Fourier transform in the directions of translational invariance,
\begin{eqnarray}
{{ \bfsfG}({\bf x},{\bf x'},\omega)}&=&\frac{1}{2\pi}\int
\mbox{d}^2{\bf k}\, e^{\imath {\bf
k}\cdot( {\bm \rho}-{\bm \rho'})}{\bfsfG}({\bf k},z,z',\omega)\nonumber\\
\end{eqnarray}
where ${\bf k}$ is a two-dimensional vector in the $x,y$-subspace and ${\bm \rho} = (x,y)$.
The Green tensor ${{\bfsfG}({\bf k},z,z',\omega)}$ assumes two different forms, depending on whether $z$ and $z'$ are
located in the same layer or not. For $z'$ in layer $j$ it is given by~\cite{Tomas1995a,Amooghorban2011a}
\begin{widetext}
\begin{subequations}\label{Green tensor at imaginary frequencies}
\begin{eqnarray}
{{\bfsfG}({\bf
k},z,z',\omega)}&=&\frac{1}{2\pi\varepsilon_0\varepsilon_j(\omega
)\omega^2}\delta{(z-z')}\hat{z}\hat{z}+\frac{{1 }}{{4\pi {\beta}_j }}\sum\limits_{\sigma  = \rm s}^{\rm p}
\xi_\sigma \frac{ { e^{ - {\beta}_j l_j } }}{{D^j_\sigma
}} \\
&&\times[{\cal{E}}_{\sigma\,>}^{(j)} ({\bf k},\omega;z){\cal{E}}_{\sigma\,<}^{(j)} (-{\bf
k},\omega;z')\Theta(z-z')+{\cal{E}}_{\sigma\,<}^{(j)} ({\bf k},\omega;z){\cal{E}}_{\sigma\,>}^{(j)} (-{\bf k},\omega;z')\Theta(z'-z) ],\,\,\,\,z\,\,\,\mbox{also in layer}\,\,j \nonumber \label{Green tensor at imaginary frequencies_a}\\
{{\bfsfG}({\bf k},z,z',\omega)}&=&\frac{{1
}}{{4\pi {\beta}_n }}\sum\limits_{\sigma  = \rm
s}^{\rm p} \xi_\sigma \frac{ t^{n/j}_\sigma e^{ - ({\beta}_j l_j
+{\beta}_n l_n )}}{D^j_\sigma
} \\
&&\times\left[{\frac{{\cal{E}}_{\sigma\,>}^{(n)} ({\bf k},\omega;z)}{D^{n/j}_{\sigma, +}}{\cal{E}}_{\sigma\,<}^{(j)}
(-{\bf k},\omega;z')\Theta(n-j)+\frac{{\cal{E}}_{\sigma\,<}^{(n)} ({\bf k},\omega;z)}{D^{n/j}_{\sigma,-}}{\cal{E}}^{\sigma\,>}_j
(-{\bf k},\omega;z')\Theta(j-n)} \right],\,\,z\,\,\, \mbox{in layer}\,
n\neq j \nonumber \label{Green tensor at imaginary frequencies_b}
\end{eqnarray}
\end{subequations}
\end{widetext}
where $\xi_{\rm s}=-1$, $\xi_{\rm p}=1$, and $\Theta(z)$ is the
usual unit step function, and
\begin{subequations}
\begin{eqnarray}
{\cal{E}}_{\sigma\,>}^{(j)} ({\bf k},\omega;z)&=&
{\bf e}_{\sigma,+}^{(j)} ({\bf q})e^{ - {\beta}_j (z - d_j)}+r_{j^ +
}^\sigma{\bf e}_{\sigma, - }^{(j)}({\bf k})e^{  {\beta}_j (z -
d_j)},\nonumber\\ \\
{\cal{E}}_{\sigma\,<}^{(j)} ({\bf k},\omega;z)&=&
{\bf e}_{\sigma, -}^{(j)}({\bf k})e^{  {\beta}_j z }+r_{j^
-}^\sigma{\bf e}_{\sigma,+}^{(j)}+({\bf k})e^{ - {\beta}_j z}.
\end{eqnarray}
\end{subequations}
Here $\sigma$ stands for ${s}-$ or ${p}$-polarization, and ${\bf e}_{{\rm s},\pm}^{(j)}=(\hat{{\bf k}}\times \hat{z})$ and ${\bf e}_{{\rm p},\pm}^{(j)}=\frac{-1}{k_j}(|{\bf k}|\hat{z}\pm{\beta}_j\hat{{\bf k}})$ are the polarization
vectors for ${s}$- and ${p}$-polarized waves propagating in the
positive-/negative-$z$ direction, with $k_j \equiv \sqrt{\omega^2
\varepsilon_j( \omega )/c^2}=k'_j+{\rm i}k''_j$  and
\begin{equation}\label{component of the wave vector}
{\beta}_j({\bf k}, \omega ) = \sqrt {\varepsilon_j( \omega )\omega^2
/c^2 -{\bf k}^2}=\beta'_j+ {\rm i} \beta''_j
\end{equation}
is the normal component of the wave vector in layer $j$'th and ${\bf k}$ the
in-plane wave vector.
Other quantities in Eqs.~(\ref{Green tensor at imaginary frequencies}) that still need to be defined are
\begin{subequations}
\begin{eqnarray}
D^{j}_\sigma  & = & 1 - r_{\sigma, -  }^{(j)}  r_{\sigma, + }^{(j)} e^{ - 2\beta_j
l_j }, \\
D^{n/j}_{\sigma, \pm} & = & 1 - r_{\sigma, \pm  }^{(n)}  r^{n\mp1/j
}_\sigma e^{ - 2\beta_n l_n },
\end{eqnarray}
\end{subequations}
where  $r_{\sigma, - }^{(j)}$ and $r_{\sigma, + }^{(j)}$ are the Fresnel
coefficients for reflection at the left/right boundary of layer $j$. In addition, $t^{n/j}_\sigma$ and $r^{n/j}_\sigma$ are the transmission and reflection coefficients between the layers $n$ and $j$.

The Green function~(\ref{Green tensor at imaginary frequencies}) is hereby defined, but not yet automatically well-defined: it is a known issue that the normal component of the wave vector Eq.~(\ref{component of the wave vector}) for active multilayer media is not automatically well defined even if the refractive index is well-defined: although the refractive index has no branch points in the upper half-plane because of causality, $\beta_j({\bf k},\omega )$ may have branch points there~\cite{Skaar2006a}. If so, then $\beta_j({\bf q},\omega )$
looses its usual physical interpretation as the propagation constant in the normal direction, since waves propagating perpendicularly to the $z$-axis propagate an infinite distance, and therefore pick up an infinite amount of gain, before arriving at any other plane $z=const$. This instability could be eliminated by limiting the extent of the active medium in the transverse direction. We will follow instead Refs.~\cite{Skaar2006a,Skaar2006b,Nistad2008a} and only consider active media without branch points where $\beta_j(\omega )$ is meaningful for real frequencies. In that case the signs of  $\beta'_j$ and $\beta''_j$ are identical to those of ${\rm Re}[\varepsilon_j(\omega )]$ and ${\rm Im}[\varepsilon_j(\omega )]$, respectively (see Refs.~\cite{Skaar2006a,Skaar2006b,Nistad2008a}).

We will now use this Green tensor~(\ref{Green tensor at imaginary frequencies}) to write the electric-field operator~(\ref{vector potential field}) in terms of creation and annihilation operators for which we will then identify input-output relations. Expressed in the same mixed Fourier representation as for the Green function, the electric-field operator becomes
\begin{eqnarray}\label{Electric field in time domain}
{\bf E}^{ (j)} ({\bf x},t)&=&\frac{1}{(2\pi)^{3/2}}\int \mbox{d}^2{\bf k}\int \mbox{d}\omega \left[{e^{{\rm i}({\bf k}\cdot {\bm \rho}-\omega t)}\,{\bf E}^{ (j)} (z,{\bf k},\omega)}\right.\nonumber\\
&&+h.c.],
\end{eqnarray}
where the electric-field component
\begin{eqnarray}\label{Electric field in frequency domain}
{\bf E}^{ (j)} (z,{\bf k},\omega)&=&\sum_{\sigma=p,s}\left[{\,{ E}^{ (j)}_{\sigma,+}(z,{\bf k},\omega){\bf e}^{(j)}_{\sigma,+}({\bf k})\nonumber}\right.\\
&&\left.{+{ E}^{ (j)}_{\sigma,-}(z,{\bf k},\omega){\bf e}^{(j)}_{\sigma,-}({\bf k})}\right],
\end{eqnarray}
is associated with light  propagation both to the right $(+)$ and left $(-)$. These `amplitudes' ${ E}^{ (j)}_{\sigma,\pm}(z,{\bf k},\omega)$  associated with right-and left propagation inside the $j$th layer can be written in terms of amplitude operators $a^{ (j)}_{\sigma,\pm}(z,{\bf k},\omega)$ as
\begin{eqnarray}\label{Electric field in term of bosonic operator}
{ E}^{ (j)}_{\sigma,\pm}(z,{\bf k},\omega)=\frac{{\rm i}\omega}{\beta_j\, c}\sqrt{\frac{\hbar \beta'_j}{2 \varepsilon_0}}\,e^{\pm{\rm i}\beta'_j z}\, a^{ (j)}_{\sigma,\pm}(z,{\bf k},\omega),
\end{eqnarray}
where the $z$-dependence  of the amplitude
operators $a^{ (j)}_{\sigma,\pm}(z,{\bf k},\omega)$ is governed by quantum Langevin equations
\begin{eqnarray}\label{right and left operators}
\frac{\partial a^{ (j)}_{\sigma,\pm}(z,{\bf k},\omega)}{\partial z}&=&\mp \beta''_j \, a^{ (j)}_{\sigma,\pm}(z,{\bf k},\omega)\\
&&\pm \frac{\sqrt{2 |\beta''_j|}}{\rm i}\, e^{\mp{\rm i}\beta'_j z} \,{f}^{(j)}_{\sigma,\pm}( z,{\bf k},\omega),\nonumber
\end{eqnarray}
so that the operators $a^{ (j)}_{\sigma,\pm}(z,{\bf k},\omega)$ and $a^{ (j)}_{\sigma,\pm}(z',{\bf k},\omega)$ for different space points within the same $j$th layer are related as
\begin{eqnarray}\label{bosonic operator for right and left}
a^{ (j)}_{\sigma,\pm}(z,{\bf k},\omega)&=&e^{\mp \beta''_j (z-z')} a^{ (j)}_{\sigma,\pm}(z',{\bf k},\omega)\,\pm\frac{\sqrt{2 |\beta''_j|}}{\rm i}\, e^{\mp \beta''_j z}\nonumber\\
&&\times\int_{z'}^z \mbox{d}z'' \, \,e^{\mp{\rm i}\beta'_j z''}\, {f}^{(j)}_{\sigma,\pm}( z'',{\bf k},\omega).
\end{eqnarray}
Here the ${f}^{(j)}_{\sigma,\pm}(\pm z',{\bf k},\omega)={\bf f}^{(j)}(\pm z',{\bf k},\omega)\cdot {\bf e}^{(j)}_{\sigma,\pm}({\bf k})$ are bosonic field operators that play the role of fundamental variables of the electromagnetic field and medium, and ${\bf f}^{(j)}(\pm z',{\bf k},\omega)$ is the partial Fourier transform of ${\bf f}({\bf x},\omega)$ in layer $j$ of Eq.~(\ref{vector potential field}). They satisfy the commutation relations
\begin{subequations}\label{commutation relations for s mode}
\begin{eqnarray}
\left[ {{f}^{(j)}_{\sigma,\pm}( z,{\bf k},\omega) , {f}^{(j')\,\dagger}_{\sigma',\pm}( z',{\bf k}',\omega')}
\right] & = & \varrho_{{\rm \sigma},+}^{(j)}\, {{\rm sgn}[\varepsilon_{{\rm
I}\,j}(\omega)]}\,\delta_{jj'}\delta_{\sigma \sigma'} \nonumber\\
&&\hspace{-2cm}\times\delta (z-z')\delta (\omega  - \omega ')\delta ({\bf k}-{\bf k}'), \\
\left[ {{f}^{(j)}_{\sigma,\pm}( z,{\bf k},\omega) , {f}^{(j')\,\dagger}_{\sigma',\mp}( z',{\bf k}',\omega')}
\right] & = & \varrho_{{\rm \sigma},-}^{(j)}\, {{\rm sgn}[\varepsilon_{{\rm
I}\,j}(\omega)]}\,\delta_{jj'}\delta_{\sigma \sigma'}\nonumber\\
&&\hspace{-2cm}\times\delta (z-z')\delta (\omega  - \omega ')\delta ({\bf k}-{\bf k}'),
\end{eqnarray}
\end{subequations}
Here, the coefficient $\rho_{\sigma,\pm}^{(j)}$ is defined to be equal to unity for $s$-polarization (i.e. for $\sigma = s$), while it is equal to $({\bf k}^2\pm|\beta_{j}|^2)/|k_{j}|^2$ for $\sigma = p$. Notice that the plus and minus subscripts in this coefficient $\rho_{\sigma,\pm}^{(j)}$ do not correspond to a propagation direction, but rather to two identical (+) and opposite (-) propagation directions.

Combining Eqs.~(\ref{Electric field in time domain})  and~(\ref{Electric field in frequency domain})  together with Eqs.~(\ref{Electric field in term of bosonic operator})  and~(\ref{bosonic operator for right and left}), the electric-field operator for the $j$th layer may be represented in a convenient form as
\begin{eqnarray}\label{vector potential field for multilayered}
{\bf E}^{ (j)} ({\bf x},t)&=&\frac{1}{2\pi}\sum_\sigma\int \mbox{d}^2{\bf k}\int_0^\infty \mbox{d}\omega\,\left[{ \frac{-{\rm i}\omega}{2\beta_j c}\sqrt{\frac{\hbar \beta'_j}{\pi \varepsilon_0}}\,e^{{\rm i}({\bf k}\cdot {\bm \rho}-\omega t)}}\right. \nonumber\\
&&\times\left({ \,e^{{\rm i}\beta'_j z}\, a^{ (j)}_{\sigma,+}(z,{\bf k},\omega){\bf e}^{(j)}_{\sigma,+}({\bf k}) }\right. \nonumber\\
&&+ \left.{\left.{e^{-{\rm i}\beta'_j z}\, a^{ (j)}_{\sigma,-}(z,{\bf k},\omega) {\bf e}^{(j)}_{\sigma,-}({\bf k}) }\right){+h.c.}}\right],
\end{eqnarray}
where the properties of the amplitude operators have now been established. It is worth mentioning that  an ordinary normal-mode expansion for the electric-field operator is recovered from this equation for frequencies $\omega$ far
from the  resonances of the medium: when gain and loss may be
disregarded, i.e. in the limit $\varepsilon_{\rm I \, j}(\omega) \rightarrow 0$, the operators $a^{ (j)}_{\sigma,\pm}(z,{\bf k},\omega)$ become mode operators independent of $z$.

These equations~(\ref{Electric field in time domain})-~(\ref{vector potential field for multilayered}) will make it possible to calculate the input and output fields at any position outside the multilayer medium recursively, without explicitly applying the multilayer Green function~(\ref{Green tensor at imaginary frequencies}).
We derive this recursive procedure in three steps: first, within each layer $j$ we relate the amplitude operators on the extreme left and right to each other, i.e. at the positions $z= z_{j-1}$ and at $z = z_{j}$. Second, we relate the operators in neighboring layers across an interface. Third, by making repeated use of the previous two steps, we can relate the amplitude operators $a^{ (1)}_{\sigma,-}(z,{\bf k},\omega)$ and $a^{ (N+1)}_{\sigma,+}(z,{\bf k},\omega)$ for the outgoing fields to the left and right of the multilayered, respectively, to the operators of the corresponding incoming fields, $a^{ (1)}_{\sigma,+}(z,{\bf k},\omega)$ and $a^{ (N+1)}_{\sigma,-}(z,{\bf k},\omega)$, and the noise operators. We discuss these three steps in some more detail below.

\sssection{Step 1} The first step is readily found by realizing that Eq.~(\ref{bosonic operator for right and left}) for $z = z_{j}$ can be written in matrix form as
\begin{eqnarray}\label{matrix relation for matrixR}
\left( {\begin{array}{*{20}c}
   {a^{ (j)}_{\sigma,+}(z_j,{\bf k},\omega)}  \\
   \\
   {a^{ (j)}_{\sigma,-}(z_j,{\bf k},\omega)}  \\
\end{array}} \right)
&=&R^{(j\,)}_\sigma
\left( {\begin{array}{*{20}c}
   {a^{ (j)}_{\sigma,+}(z_{j-1},{\bf k},\omega)}  \\
   \\
   {a^{ (j)}_{\sigma,-}(z_{j-1},{\bf k},\omega)}  \\
\end{array}} \right)\nonumber\\
&&+
\left( {\begin{array}{*{20}c}
   {c^{ (j)}_{\sigma ,+}({\bf k},\omega)}  \\
   \\
   {c^{ (j)}_{\sigma,-}({\bf k},\omega)}  \\
\end{array}} \right)
\end{eqnarray}
where $R^{(j\,)}_\sigma$ is a diagonal $2\times 2$ matrix with
$R^{(j\,)}_{\sigma,11}=1/R^{(j\,)}_{\sigma,22}=e^{-\beta''_{j}l_j}$.
The quantum noise operators in this matrix equation~(\ref{matrix relation for matrixR}) are given by
\begin{eqnarray}\label{elements of Matrixc}
c^{ (j)}_{\sigma,\pm}({\bf k},\omega)&=&\pm\frac{\sqrt{2 |\beta''_j|}}{\rm i}\, e^{\mp \beta''_j z_j} \nonumber\\
&&\times\int_{z_{j-1}}^{z_{j}} \mbox{d}z' \, \,e^{\mp{\rm i}\beta'_j z'}\, { f}^{(j)}_{\sigma,\pm}(z',{\bf k},\omega),
\end{eqnarray}
and evidently these inhomogeneous terms in the matrix relation~(\ref{matrix relation for matrixR})  are the qualitative novelty as compared to the standard transfer-matrix analysis of multilayer media in classical electrodynamics.
Recalling the commutation relations~(\ref{commutation relations for s mode}), the operators $c^{ (j)}_{\sigma,\pm}({\bf k},\omega)$ are found to satisfy the commutation relations
\begin{subequations}
\begin{eqnarray}
\left[ {c^{ (j)}_{\sigma,\pm}( {\bf k},\omega) , c^{ (j)\,\dagger}_{\sigma',\pm}( {\bf k}',\omega')}
\right] & = & \,2\,\rho_{\sigma,+}^{(j)}\, e^{\mp\beta''_{j}l_{j}}\sinh({\beta''_{j}}l_{j})\,\delta_{\sigma \sigma'}\nonumber\\
&&\hspace{-2cm}\times\delta (\omega  - \omega ')\delta^{2} ({\bf k}-{\bf k}'),   \\
\left[ {c^{ (j)}_{\sigma,\pm}( {\bf k},\omega) , c^{ (j)\,\dagger}_{\sigma',\mp}( {\bf k}',\omega')}
\right] & = & \,-\frac{2\beta''_{j}}{\beta'_{j}}\,\rho_{\sigma,-}^{(j)}\, e^{\mp {\rm i}\beta'_{j}(z_{j}+z_{j-1})}\nonumber\\
&&\hspace{-2cm}\times\sin(\beta'_{j}l_{j})\,\delta_{\sigma \sigma'}\delta (\omega  - \omega ')\delta^{2} ({\bf k}-{\bf k}').
\end{eqnarray}
\end{subequations}

\sssection{Step 2} In the second step, we relate the operators $a^{ (j+1)}_{\sigma,\pm}(z_j,{\bf k},\omega)$ and
$a^{ (j)}_{\sigma,\pm}(z_j,{\bf k},\omega)$ in neighboring layers across the interface at $z_{j}$ to each other by using the form Eq.~(\ref{Green tensor at imaginary frequencies_b}) for the Green function ${{\bfsfG}({\bf k},z,z',\omega)}$ for positions $z,z'$ in neighboring layers. This Green function already by construction respects the Maxwell boundary conditions that  the  tangential components of the electric and magnetic fields be continuous. We obtain the operator matrix relation
\begin{equation}\label{matrix relation for matrixS}
\left( {\begin{array}{*{20}c}
   {a^{ (j+1)}_{\sigma,+}(z_j,{\bf k},\omega)}  \\
   \\
   {a^{ (j+1)}_{\sigma,-}(z_j,{\bf k},\omega)}  \\
\end{array}} \right)=S^{(j)}_{q}
\left( {\begin{array}{*{20}c}
   {a^{ (j)}_{\sigma,+}(z_j,{\bf k},\omega)}  \\
   \\
   {a^{ (j)}_{\sigma,-}(z_j,{\bf k},\omega)}  \\
\end{array}} \right),
\end{equation}
which also holds for classical amplitudes and where the matrix $S^{(j)}_{\sigma}$ is given by
\begin{widetext}
\begin{equation}\label{matrix relation for matrixSs}
S^{(j)}_{\sigma}=\frac{1}{2\beta_j}\sqrt{\frac{\beta'_j}{\beta'_{j+1}}}\left( {\begin{array}{*{20}c}
   {(\beta_{j+1}\kappa_{\sigma,j/j+1}+\beta_{j}\kappa_{\sigma,j+1/j})e^{{\rm i}(\beta'_{j}-\beta'_{j+1})z_{j}}} \,\,\,\,\,\,\,{(\beta_{j+1}\kappa_{\sigma,j/j+1}-\beta_{j}\kappa_{\sigma,j+1/j})e^{-{\rm i}(\beta'_{j}+\beta'_{j+1})z_{j}}}  \\
   \\
   {(\beta_{j+1}\kappa_{\sigma,j/j+1}-\beta_{j}\kappa_{\sigma,j+1/j})e^{{\rm i}(\beta'_{j}+\beta'_{j+1})z_{j}}} \,\,\,\,\,\,\,{(\beta_{j+1}\kappa_{\sigma,j/j+1}+\beta_{j}\kappa_{\sigma,j+1/j})e^{-{\rm i}(\beta'_{j}-\beta'_{j+1})z_{j}}}  \\
\end{array}} \right),
\end{equation}
\end{widetext}
in which $\kappa_{s,j/j+1}=1$ and $\kappa_{p,j/j+1}=k_j/k_{j-1}$.

\sssection{Step 3} In the third, final step, we invoke Eqs.~(\ref{matrix relation for matrixR}) and~(\ref{matrix relation for matrixS}) alternatingly and repeatedly, until we finally obtain the operator of the outgoing fields to the leftmost and rightmost layers, respectively, $a^{(1)}_{\sigma,-}(z_1,{\bf k},\omega)$ and $a^{(N+1)}_{\sigma,+}(z_{\rm N},{\bf k},\omega)$, in terms of the two incoming fields $a^{(1)}_{\sigma,+}(z_1,{\bf k},\omega)$ and
$a^{(N+1)}_{\sigma,-}(z_{\rm N},{\bf k},\omega)$, as well as the noise fields. The sought input-output relation for amplitude operators is thereby obtained as
\begin{equation}\label{matrix relation for output input}
\left( {\begin{array}{*{20}c}
   {a^{(1)}_{\sigma,-}(z_1)
   %{\bf k},\omega)
   }  \\
   \\
   {a^{(N+1)}_{\sigma,+}(z_{\rm N})
   %,{\bf k},\omega)
   }  \\
\end{array}} \right)={\cal A}_{\sigma}
\left( {\begin{array}{*{20}c}
   {a^{(1)}_{\sigma,+}(z_1)
   %,{\bf k},\omega)
   }  \\
   \\
   {a^{(N+1)}_{\sigma,-}(z_{\rm N})
   %,{\bf k},\omega)
   } \\
\end{array}} \right)+\left( {\begin{array}{*{20}c}
   {F_{\sigma,-}
   %({\bf k},\omega)
   }  \\
   \\
   {F_{\sigma,+}%({\bf k},\omega)
   } \\
\end{array}} \right),
\end{equation}
where we suppressed the $({\bf k},\omega)$-dependence, and where the quantum noise originating from all layers with either loss or gain is given by
\begin{equation}\label{definition of F operators}
\left( {\begin{array}{*{20}c}
   {F_{\sigma,-}
   %({\bf k},\omega)
   }  \\
   \\
   {F_{\sigma,+}%({\bf k},\omega)
   } \\
\end{array}} \right)=
{\cal B}^{(2)}_{\sigma}
\left( {\begin{array}{*{20}c}
   {c^{(2)}_{\sigma,+}
   %%({\bf k},\omega)
   }  \\
   \\
   {c^{(2)}_{\sigma,-}
   %({\bf k},\omega)
   } \\
\end{array}} \right)+\cdots+{\cal B}^{(N)}_{\sigma}
\left( {\begin{array}{*{20}c}
   {c^{(N)}_{\sigma,+}
   %({\bf k},\omega)
   }  \\
   \\
   {c^{(N)}_{\sigma,-}
   %({\bf k},\omega)
   } \\
\end{array}} \right),
\end{equation}
in which the coefficient matrices ${\cal A}_{\sigma}$ and ${\cal B}_{\sigma}$ are given by
\begin{subequations}
\begin{equation}\label{definition of matrix A}
{\cal A}_{\sigma}=A_{\sigma\,22}^{-1}\left( {\begin{array}{*{20}c}
   {-A_{\sigma 21}} \,\,& {1}  \\
   {A_{\sigma 11}\,A_{\sigma 22}-A_{q 12}\,A_{\sigma 21}} \,\,& {A_{\sigma 12}}  \\
\end{array}} \right),
\end{equation}
\begin{eqnarray}\label{definition of matrix B}
&&{\cal B}^{(j\,)}_{\sigma }=A_{\sigma \,22}^{-1}\\
&&\times\left( {\begin{array}{*{20}c}
   {-B_{\sigma  21}^{(j\,)}} \,\,& {-B_{\sigma  22}^{(j\,)}}  \\
   {B_{\sigma  11}^{(j\,)}\,A_{\sigma  22}-A_{\sigma  12}\,B_{\sigma  21}^{(j\,)}} \,\,& {B_{\sigma  12}^{(j\,)}\,A_{\sigma  22}-A_{\sigma  12}\,B_{\sigma  22}^{(j\,)}}  \\
\end{array}} \right).\nonumber
\end{eqnarray}
\end{subequations}
Here, the matrices $B^{(j\,)}_{\sigma }$ satisfy the recursion relations
$B^{(k-1\,)}_{\sigma }=B^{(k\,)}_{\sigma }\cdot R^{(k\,)}_{\sigma } \cdot S^{(k-1\,)}_{\sigma }$ and
$B^{(N\,)}_{\sigma }=S^{(N\,)}_{\sigma }$, with $k=3,4,\ldots,N$, and $A_{\sigma }=B^{(2\,)}_{\sigma }\cdot
R^{(2\,)}_{\sigma }\cdot S^{(1\,)}_{\sigma }$.
From Eq.~(\ref{matrix relation for output input}) we can appreciate that the multiple transmissions and reflections in the multilayer medium of the incident light are described by the same transfer matrices ${\cal A}_\sigma$ as in classical optics, whereas the
matrix elements ${\cal B}^{(j\,)}_\sigma$ have no classical analogues, since they  describe the   propagation of quantum noise that originates from layer $j$.

\sssection{Input and output outside the multilayer}
In this article we will mostly focus on the transmitted and reflected output states of light as compared to the optical input states. Let us therefore specify how from the above general formalism the field operators outside of the multilayer structure are obtained.
The amplitude operators of the incoming fields $a^{ (1)}_{\sigma,+}(z,{\bf k},\omega)$ and $a^{ (N+1)}_{\sigma,-}(z,{\bf k},\omega)$  in Eq.~(\ref{matrix relation for output input}) are defined within the space intervals $-\infty< z \leq z_1$ and $z_{N+1} \leq z < \infty$, respectively, see the sketch in Fig.~\ref{Fig:Scheme of the multilayer dielectric}. The explicit form of these input operators can be obtained with the use of Eq.~(\ref{bosonic operator for right and left})  as
\begin{subequations}\label{bosonic_input_operators}
\begin{eqnarray}
a^{ (1)}_{\sigma,+}(z,{\bf k},\omega)
& =&
\frac{1}{\rm i}\sqrt{2 |\beta''_1|}\,\,e^{-\beta''_1 z}\\
&&\times\int_{-\infty}^{ z} \mbox{d}z'\, e^{-{\rm i}\beta_1 z'}\,{ f}^{(1)}_{\sigma,+}( z',{\bf k},\omega),\nonumber
\label{bosonic operator for domain 1} \\
a^{ (N+1)}_{\sigma,-}(z,{\bf k},\omega)
&=&
\frac{1}{\rm i}\sqrt{2 |\beta''_{N+1}|}\,\,e^{\beta''_{N+1} z}\\
&&\times\int_{z}^{\infty} \mbox{d}z'\, e^{{\rm i}\beta_{N+1} z'}\,{ f}^{(N+1)}_{\sigma,-}( z',{\bf k},\omega).\nonumber
\label{bosonic operator for domain N+1}
\end{eqnarray}
\end{subequations}
Using the above relations~(\ref{bosonic_input_operators}) and the commutation relation~(\ref{commutation relations for s mode}), we find that the input operators satisfy the commutation relations
\begin{subequations}
\begin{eqnarray}
\left[a^{ (1)}_{\sigma,+}(z,{\bf k},\omega), a^{ (1)\, \dag}_{\sigma',+}(z',{\bf k'},\omega')\right]
&=&
\varrho_{{\rm \sigma},+}^{(1)}\,e^{-\beta''|z-z'|}\nonumber\\
&&\hspace{-3.5cm} \times{\rm sgn}[\varepsilon_{{\rm
I}\,1}(\omega)]\,\delta_{\sigma\sigma'}\delta (\omega  - \omega ')\delta ({\bf k}-{\bf k}'),
\label{commutation relation for operator a in domain 1}
\end{eqnarray}
and an analogous relation holds  for the operators in the other outer layer labeled $N+1$ on the opposite size of the metamaterial.
It also follows that input operators  of different outer layers commute,
\begin{equation}  \label{commutation relation for operator a in domain 1 and N+1}
\left[a^{ (1)}_{\sigma,+}(z,{\bf k},\omega), a^{ (N+1)\, \dag}_{\sigma',-}(z',{\bf k'},\omega')\right]
=
0,
\end{equation}
\end{subequations}
as one would expect for these independent input channels.

The commutation relations for the output amplitude operators
$a^{ (1)}_{\sigma,-}(z,{\bf k},\omega)$ and $a^{ (N+1)}_{\sigma,+}(z,{\bf k},\omega)$ can also be determined. We do not spell them out here, but they can be
derived by applying the input-output relation~(\ref{matrix relation for output input}) and the commutation relation~(\ref{commutation relation for operator a in domain 1}).
Indeed, the input-output relation~(\ref{matrix relation for output input}) together with the commutation relations~(\ref{commutation relation for operator a in domain 1})--(\ref{commutation relation for operator a in domain 1 and N+1}), contain all
information necessary to transform an arbitrary function
of the input-field operators into the corresponding function
of the output-field operators. In particular, it enables
one to express arbitrary moments and correlations
of the outgoing fields in terms of those of the incoming fields
and the quantum-noise excitations in the multilayers. For normal incidence (${\bf k}=0$), this general input-output relation reduces to the well-known relation given in~\cite{Gruner1996a}, which we also made use of in our Ref.~\onlinecite{Amooghorban2013a}. The somewhat lengthy equations for general multilayers of the present section reduce to a much simpler form when specified for a single layer, as will be done below in Sec.~\ref{Sec:effective_index}.

\section{Quantum optical effective-index theory}\label{Sec:effective_index}
%%%%%%%%%%%%%%%%%%%%%%%%%%%%%%%%%%%%%%%%%%%%%%%%%%%%%%%%%%%%%%%%%%%%%

In  Sec.~\ref{Sec:Multilayer} we presented an input-output formalism for the electromagnetic field operators in arbitrary layered media with gain and loss, including quantum noise terms. So in principle, the problem is solved how output fields depend on the input, also for layered metamaterials: periodic multilayer media with unit cells much smaller than optical wavelengths. However, for these layered metamaterials one can hope that a simpler, effective description as a homogeneous medium is also possible, here in quantum optics just like it is known to be possible in classical optics.

Expressing quantum noise in terms of the effective index presupposes that we know how to determine the effective material parameters of a metamaterial. This has been thoroughly studied (and still is an active field of study) in classical optics~\cite{Smith2002a,Smith2006a,Acher2007a,Sun2009a,Felbacq2009a,Andryieuski2009a,Mortensen2010a}. The effective index in quantum optics is the same as in classical optics and can be determined using the same methods. We will use and compare two such methods. First, the scattering method by Smith and co-workers~\cite{Smith2002a,Smith2006a,Mortensen2010a} boils down to finding the effective index of a homogeneous medium that mimicks best the transmission and reflection off the metmaterial. Second, we use the dispersion method, where effective parameters are obtained from the small-$(k,\omega)$ Taylor expansion of the known dispersion relation of periodic multilayer structures.  For completeness we briefly present both methods in Appendix~\ref{App:A}.
We will focus on metamaterials with strongly subwavelength unit cells, for which unique effective parameters can be identified, practically independent of the method used to obtain them.

In this section we present a quantum optical effective-index theory. By this we mean an effective-medium theory that describes the metamaterial entirely in terms of its effective index (or equivalently, in terms of its effective dielectric function). In that sense it does not differ from the usual effective theory in classical optics. But it differs from the usual effective-index theories in classical optics because quantum noise is also described. The crucial assumption is that also the quantum noise of the metamaterial can be described solely in terms of its effective dielectric function.
The effective-index theory presented in this section concerns three-dimensional light propagation in layered metamaterials, thereby generalizing the effective-index theory of Ref.~\cite{Amooghorban2013a} to arbitrary propagation directions and for two distinct types of polarization.
The accuracy of the effective-index theory will first be tested in calculations of output intensities in Sec.~\ref{sec:intensity}.

%%%%%%%%%%%%%%%%%%%%%%%%%%%%%%%%%%%%%%%%%%%
\sssection{Output operators of a single homogeneous layer}
Assume that we have used either the scattering or the dispersion method to determine the values for the effective dielectric tensor components for our multilayer structure. And assume that in classical optics the entire structure can effectively be described as a single dielectric layer. Then we can also try and apply the elaborate quantum optical input-output formalism of Sec.~\ref{Sec:Multilayer} to that single effective layer.  With the two planar interfaces of the homogenized slab located at $z_1=0$ and $z_{\rm N}={\rm L}$, the input-output relation~(\ref{matrix relation for output input}) for the single effective layer reduces to the simpler form
\begin{eqnarray}\label{matrix relation for output input and effective medium}
\left( {\begin{array}{*{20}c}
   {a^{(1)}_{\sigma,-}(z_1,{\bf k},\omega)}  \\
   \\
   {a^{(N+1)}_{\sigma,+}(z_{\rm N},{\bf k},\omega)}  \\
\end{array}} \right)&=& {\cal A}_{{\rm eff},\,\sigma}
\left( {\begin{array}{*{20}c}
   {a^{(1)}_{\sigma,+}(z_1,{\bf k},\omega)}  \\
   \\
   {a^{(N+1)}_{\sigma,-}(z_{\rm N},{\bf k},\omega)} \\
\end{array}} \right)\nonumber \\
&&+\left( {\begin{array}{*{20}c}
   {F_{{\rm eff}\,\sigma,-}({\bf k},\omega)}  \\
   \\
   {F_{{\rm eff}\,\sigma,+}({\bf k},\omega)} \\
\end{array}} \right),
\end{eqnarray}
where according to Eq.~(\ref{definition of matrix A}), the matrix presentation ${\cal A}_{{\rm eff},\,\sigma}$ is equal to
\begin{equation}\label{eq:Aeffdef}
\left( {\begin{array}{*{20}c}
   {r_{{\rm eff,\,\sigma}}\,\,\,\,} & {t_{\rm eff,\,\sigma}}  \\
   \\
   {t_{\rm eff,\,\sigma}\,\,\,\,} & {e^{-2{\rm i}\beta_0 d}\, r_{\rm eff,\,\sigma}}  \\
\end{array}} \right),
\end{equation}
and where the effective complex reflection and transmission amplitudes of the homogenized slab are given by the well-known classical expressions
\begin{subequations}\label{Eq:classical_r_and_t}
\begin{eqnarray}
r_{\rm eff,\,\sigma}&=&\frac{(\beta_{\rm eff,\sigma}^2-\beta_{0}^2)\left({\exp{[2{\rm i}\beta_{\rm eff}{\rm L}]}-1}\right)}{(\beta_{\rm eff,\sigma}+\beta_{0})^2-(\beta_{\rm eff,\sigma}-\beta_{0})^2\exp{[2{\rm i}\beta_{\rm eff}{\rm L}]}} \label{R effsp} \\
t_{\rm eff,\,\sigma}&=&\frac{4\beta_{\rm eff,\sigma}\beta_{0}\exp{[{\rm i}(\beta_{\rm eff}-\beta_{0}){\rm L}]}}{(\beta_{\rm eff}+\beta_{0,\sigma})^2-(\beta_{\rm eff,\sigma}-\beta_{0})^2\exp{[2{\rm i}\beta_{\rm eff}{\rm L}]}}.\label{T effsp}\,\,\,\,\,\,\,\,\,\,\,\,
\end{eqnarray}
\end{subequations}
Here, $\beta_{{\rm eff},\sigma}$ stands for  $\beta_{{\rm eff},s}=\beta_{\rm eff}$ and $\beta_{{\rm eff},p}=\beta_{\rm eff}/\varepsilon_{\rm eff}$. The effective noise operator $F_{{\rm eff},\,\sigma}$ has no classical analogue. It represents the quantum noise associated with loss and gain and combinations thereof inside this effective medium, and in the present effective-index theory its right- (+) and left-going (--) components are given by
\begin{widetext}
\begin{subequations}\label{Neffsp}
\begin{eqnarray}\label{Neffsp-}
F_{{\rm eff,\,\sigma}-}({\bf k},\omega)&=&\frac{-2{\rm i}\beta_{0}\sqrt{2\beta'_{\rm eff,\sigma}\beta''_{\rm eff}/\beta'_{0}}}{(\beta_{\rm eff,\sigma}+\beta_{0})^2-(\beta_{\rm eff,\sigma}-\beta_{0})^2\exp{[2{\rm i}\beta_{\rm eff}{\rm L}]}}\left({(\beta_{\rm eff,\sigma}-\beta_{0})e^{2{\rm i}\beta_{\rm eff}d}\int_0^{{\rm L}} dz' e^{-{\rm i}\beta_{\rm eff}z'}f_{\rm eff\, \sigma,+}(z',{\bf k},\omega)}\right.\nonumber\\
&& + \left.{(\beta_{\rm eff,\sigma}+\beta_{0})\int_0^{{\rm L}} dz' e^{{\rm i}\beta_{\rm eff}\omega z'}f_{\rm eff\, \sigma,-}(z',{\bf k},\omega)}\right),
\end{eqnarray}
\begin{eqnarray}\label{Neffsp+}
F_{{\rm eff,\,\sigma}+}({\bf k},\omega)&=&\frac{-2{\rm i}\beta_{0}\sqrt{2\beta'_{\rm eff,\sigma}\beta''_{\rm eff}/\beta'_{0}}\exp{[{\rm i}(\beta_{\rm eff}-\beta_{0}){\rm L}]}}{(\beta_{\rm eff,\sigma}+\beta_{0})^2-(\beta_{\rm eff,\sigma}-\beta_{0})^2\exp{[2{\rm i}\beta_{\rm eff}{\rm L}]}}\left({(\beta_{\rm eff,\sigma}+\beta_{0})\int_0^{{\rm L}} dz' e^{-{\rm i}\beta_{\rm eff}z'}f_{\rm eff\, \sigma,+}(z',{\bf k},\omega)}\right.\nonumber\\
&&+ \left.{(\beta_{\rm eff,\sigma}-\beta_{0})\int_0^{{\rm L}} dz' e^{{\rm i}\beta_{\rm eff}\omega z'}f_{\rm eff\, \sigma,-}(z',{\bf k},\omega)}\right),
\end{eqnarray}
\end{subequations}
\end{widetext}
with the commutation relations
\begin{subequations}\label{commutation_Feff}
\begin{eqnarray}
&&[ F_{{\rm eff\, \sigma},\pm}({\bf k},\omega), \, F^{\dag}_{{\rm eff\,\sigma'}, \pm}({\bf k'},\omega')]  =  \\
&& (1-|r_{\rm eff,\,\sigma}|^2-|t_{\rm eff,\,\sigma}|^2) \,\delta_{\sigma\,\sigma'} \delta({\bf k}-{\bf k'}) \delta(\omega-\omega'), \nonumber \label{commutationFeffplusplus}
\end{eqnarray}
\begin{eqnarray}
&&  [ F_{{\rm eff\, \sigma},\pm}({\bf k},\omega), \, F^{\dag}_{{\rm eff\,\sigma'}, \mp}({\bf k'},\omega')]  =  \\
&& -\left({r_{\rm eff,\,\sigma}t^*_{\rm eff,\,\sigma}+e^{2{\rm i}\beta_0 d} \, r^*_{\rm eff,\,\sigma}t_{\rm eff,\,\sigma}}\right) \,\delta_{\sigma\,\sigma'} \delta({\bf k}-{\bf k'}) \delta(\omega-\omega'), \nonumber \label{commutationFeffplusminus}
\end{eqnarray}
\end{subequations}
in terms of the classical amplitude reflection and transmission amplitudes of Eq.~(\ref{Eq:classical_r_and_t}).
Furthermore, just like for the general multilayer in Sec.~\ref{Sec:Multilayer}, the optical input operators of the effective slab satisfy the bosonic commutation relations
\begin{eqnarray}\label{commutation relation for input operator for single slab}
[{a^{(1)}_{\sigma,+}({\bf k},\omega)},\,{a^{(1)\,\dag}_{\sigma',+}({\bf k}',\omega')}]&=&[{a^{(N+1)}_{\sigma,-}({\bf k},\omega)},\,{a^{(N+1)\,\dag}_{\sigma',-}({\bf k}',\omega')}]\nonumber \\
&=&\delta_{\sigma\,\sigma'}\delta({\bf k}-{\bf k'})\delta(\omega-\omega'),
\end{eqnarray}
since the input operators in free space incident on the slab cannot sense the presence of the effective slab before these input waves arrive at it. It was therefore to be expected (and a consistency test) that the input operators turned out to have the same commutators as the corresponding quantum operators in free space.

Using the previous two commutation relations and the input-output relations~(\ref{matrix relation for output input and effective medium}), the bosonic commutation relations for the output-mode operators $a^{(1)}_{\sigma,-}$ and $a^{(N+1)}_{\sigma,+}$ can then also be obtained,
\begin{eqnarray}\label{commutation relation for output operator for single slab}
[{a^{(1)}_{\sigma,-}({\bf k},\omega)},\,{a^{(1)\,\dag}_{\sigma',-}({\bf k}',\omega')}]&=&[{a^{(N+1)}_{\sigma,+}({\bf k},\omega)},\,{a^{(N+1)\,\dag}_{\sigma',+}({\bf k}',\omega')}] \nonumber \\
& =& \delta_{\sigma\,\sigma'}\delta({\bf k}-{\bf k'})\delta(\omega-\omega').
\end{eqnarray}
The quantum optical effective-index theory for planar metamaterials is hereby defined.
Based on these expressions for the effective output operators and their commutators, we will in Sec.~\ref{sec:intensity} compare predictions for physical observables using the quantum optical effective-index theory as compared to the full multilayer quantum theory of the previous section.

\section{First test: power spectra}\label{sec:intensity}

%%%%%%%%%%%%%%%%%%%%%%%%%%%%%%%%%%%%%%%%%%%%%%%%%%%%%%%%%%%%%%%%%%%%%
%
As a first test and comparison of the quantum optical effective-index theory, we will now study the output intensities of light due to spontaneously emitted photons. If atoms that make up the metamaterial are excited, either thermally or because of external pumping, then they can decay spontaneously. This is a known noise source in lasers, which is typically overlooked for metamaterials.
There is a variety of different quantum definitions of the power spectrum in the literature~\cite{Cresser1983}.
Here we choose the quantum generalization of the classical definition of the energy spectrum for the case of a stationary field~\cite{Cresser1983}. Just like its classical counterpart, it is directly related to observables in light detection experiments. For sufficiently small pass-band width of the spectral apparatus, the power spectrum $\mathscr{S}({\bf x},\omega)$ of the light emitted on the right-hand side of our multilayer metamaterial of Fig.~\ref{Fig:Scheme of the multilayer dielectric}  is given by
\begin{widetext}
\begin{eqnarray}\label{definition energy spectrum}
{\mathscr{S}}({\bf x},\omega)&=&\underset{T\to \infty }{\mathop{\lim }}\frac{1}{2\pi T} \iint_{-T/2}^{T/2} \mbox{d}t \mbox{d}t'\,e^{-{\rm i}\omega(t-t')}\langle {\bf E}^{(N+1)-}({\bold x},t)\cdot {\bf E}^{(N+1)+}({\bold x},t') \rangle,
\end{eqnarray}
\end{widetext}
where $\omega$ is the operating frequency of the spectral apparatus, and $T$ is the duration the detector is switched on. Here, the positive-frequency part of the electric field operator ${\bf E}^{N+1(+)}$  can now be determined with the help of Eq.~(\ref{Electric field in frequency domain}). As usual the negative-frequency part of the field is obtained by taking the Hermitian conjugate of Eq.~(\ref{Electric field in term of bosonic operator}). We will also need  the  input-output relation~(\ref{matrix relation for output input}) to express the  annihilation and creation operators of the outgoing field on the right-hand side of the  multilayer in terms of the operators of the ingoing fields and of the quantum noise. The power spectrum~(\ref{definition energy spectrum}) can then be written as
\begin{widetext}
\begin{eqnarray}\label{power spectrum for loss-compensation multylayer}
{\mathscr{S}}(z,\omega)&=& \frac{\hbar\omega^2}{2\,\varepsilon_0\,c^2 } \sum_{\sigma}   \int \mbox{d} {\bf k}\,\beta_0^{-1}\left\{{
\langle a^{(N+1)\,\dag}_{\sigma,-}(z,{\bf k},\omega)\, a^{(N+1)}_{\sigma,-}(z,{\bf k},\omega) \rangle\,\left({ |{\cal A}_{\sigma, 22}|^2 + e^{-2{\rm i} \beta_0 z}{\cal A}_{\sigma, 22}^* \,\varrho_{\sigma,-}^{(N+1)}}\right. }\right.\nonumber\\
&&\hspace{8.7cm}\left.{+ e^{2{\rm i}\beta_0 z}{\cal A}_{\sigma, 22}\, \varrho_{-}^{(N+1)}+1}\right)\nonumber\\
&&\hspace{3.5cm} +\langle a^{(N+1)\,\dag}_{\sigma,-}(z,{\bf k},\omega)\, a^{(1)}_{\sigma,+}(z,{\bf k},\omega) \rangle\,\left({ {\cal A}_{\sigma, 21} \,{\cal A}_{\sigma, 22}^* \,+ e^{2{\rm i} \beta_0 z}{\cal A}_{\sigma, 21}\, \varrho_{\sigma\,-}^{(N+1)}}\right)\nonumber\\
&& \hspace{3.5cm}+\langle a^{(1)\,\dag}_{\sigma,+}(z,{\bf k},\omega)\, a^{(N+1)}_{\sigma,-}(z,{\bf k},\omega) \rangle\,\left({ {\cal A}_{\sigma, 21}^* \,{\cal A}_{\sigma, 22}\, + e^{-2{\rm i} \beta_0 z}{\cal A}_{\sigma, 21}^* \,\varrho_{\sigma\,-}^{(N+1)}}\right)\nonumber\\
&&\hspace{3.5cm}\left.{ +\langle a^{(1)\,\dag}_{\sigma,+}(z,{\bf k},\omega)\, a^{(1)}_{\sigma,+}(z,{\bf k},\omega) \rangle\, |{\cal A}_{\sigma, 21}|^2 }\right\} +{\mathscr{S}}_{\rm Spon}(\omega),
\end{eqnarray}
\end{widetext}
where the parameters $\varrho_{\sigma\,-}^{(N+1)}$ equal to unity and  $(2k^2\,c^2/\omega^2-1)$ for $s$-and $p$-polarized light, respectively, so they coincide for normal incidence as they should. Expectation values are denoted by brackets. Thanks to the input-output  theory, the brackets occur around products of input operators and  all expectation values are taken with respect to both the states of the incoming optical fields and the states of the noise fields within all medium layers.
All terms except the last one on the right-hand side of Eq.~(\ref{power spectrum for loss-compensation multylayer})
describe output photons caused by (multiple) reflections and transmissions of input photons.
The final term ${\mathscr{S}}_{\rm Spon}({\bf x},\omega)$ on the other hand is independent of the optical input signal, and is determined by the quantum noise in the medium, especially by the properties of the noise operators defined in Eq.~(\ref{definition of F operators}), as detailed below. Physically, thermal excitations in passive layers and especially pumped excitations in amplifying layers give rise to spontaneously emitted noise photons.

We want to know how well quantum optical effective-medium theories describe the amount of quantum noise photons that contribute to photon-counting measurements.  In this section we will therefore study output intensities in the absence of any optical input signal, in other words all optical incoming fields are assumed to be in the vacuum state  $|0 \rangle$. In that simple case all terms except the last one in Eq.~(\ref{power spectrum for loss-compensation multylayer}) vanish identically and all output photons are spontaneously emitted noise photons, or ${\mathscr{S}}({\bf x},\omega)={\mathscr{S}}_{\rm Spon}({\bf x},\omega)$.
We will now use our input-output formalism once more, this time to express the spontaneously emitted light in terms of the noise sources in the multilayer medium. In particular, using the matrix form~(\ref{definition of matrix B}) of the factors ${\cal B}^{(j\,)}$, the power spectrum~(\ref{power spectrum for loss-compensation multylayer}) becomes
\begin{eqnarray}\label{Power spectrum of of spontaneous emission}
{\mathscr{S}}_{\rm Spon}(\omega)
& =& \sum_{\sigma}\int_{0}^{\pi/2} \mbox{d}\theta\, {\mathscr{S}}_{\rm Spon,\sigma}(\theta,\omega)  \\
&=&\frac{\hbar\omega^2}{8\pi^2 \varepsilon_0 c^2}\sum_{\sigma}\int \mbox{d}{\bf k}\,\beta_0^{-1} \, \langle F^\dag_{{ \sigma},+}({\bf k},\omega) \,F_{{\sigma}, +}({\bf k},\omega)\rangle  \nonumber
\end{eqnarray}
from which it is clear that indeed  the power spectrum of the spontaneously emitted light depends on the quantum
noise through the expectation value of $\langle F^\dag_{{ \sigma},+}({\bf k},\omega) \,F_{{\sigma}, +}({\bf k},\omega)\rangle$.
%
%%%%%%%%%%%%%%%%%%%%%%%%%%%%%%%%%%%%%%%%%%%%%%%%%%%%%%%%%%%%%%%%%%%%%%%%%
%
\begin{figure*}[t]
%\begin{minipage}[b]{0.495\linewidth}
%\begin{minipage}[b]{0.8\linewidth}
%centering
%\includegraphics[width=\textwidth]{IS30.eps}
%\includegraphics[width=\textwidth]{IS30.png}
\includegraphics[width = \columnwidth]{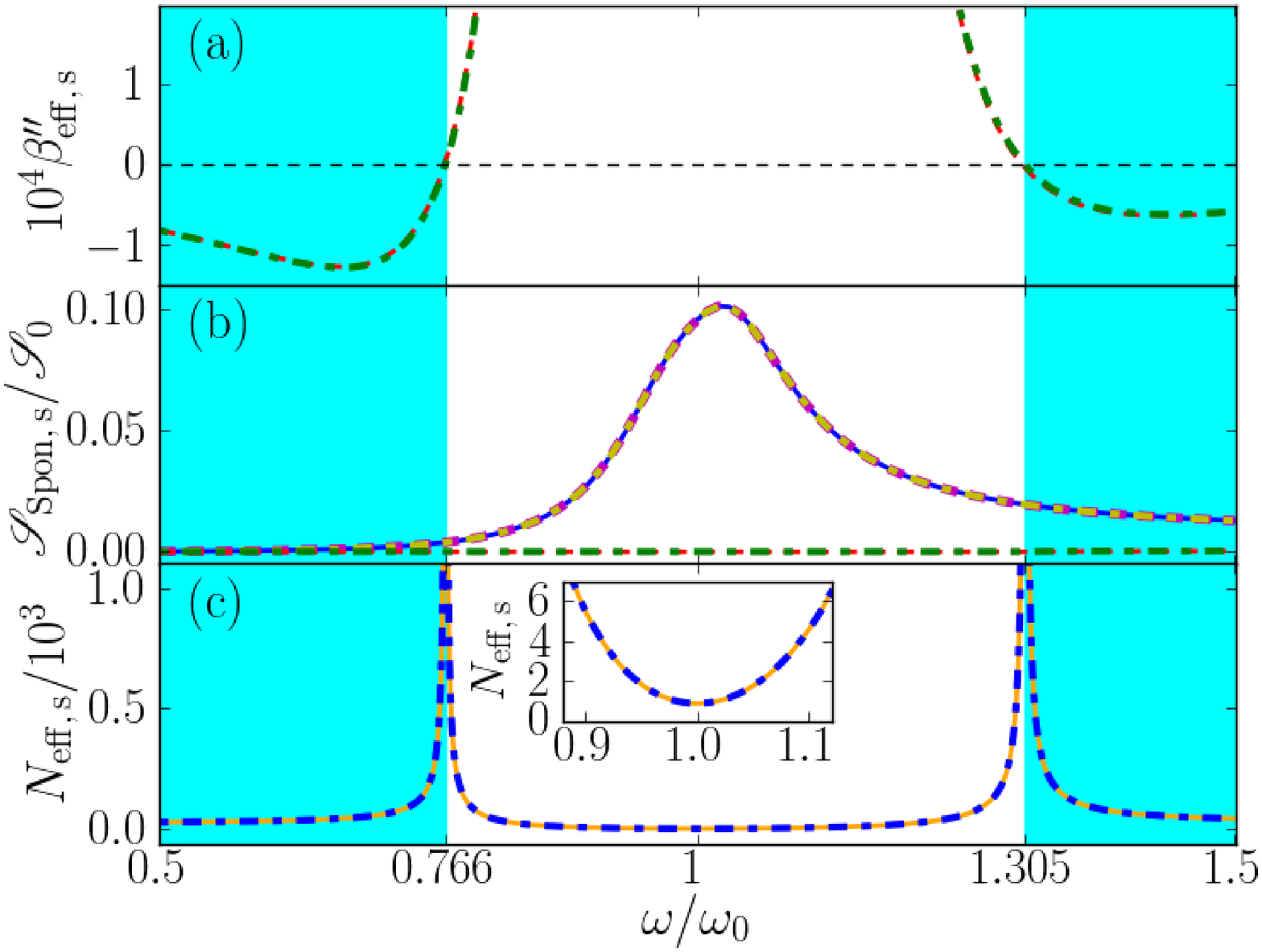}
%\end{minipage}
%\hspace{0cm}
%begin{minipage}[b]{0.49\linewidth}
%\begin{minipage}[b]{0.8\linewidth}
%\centering
%\includegraphics[width=\textwidth]{IP30.eps}
%\includegraphics[width=\textwidth]{IP30.png}
%\end{minipage}
\includegraphics[width = \columnwidth]{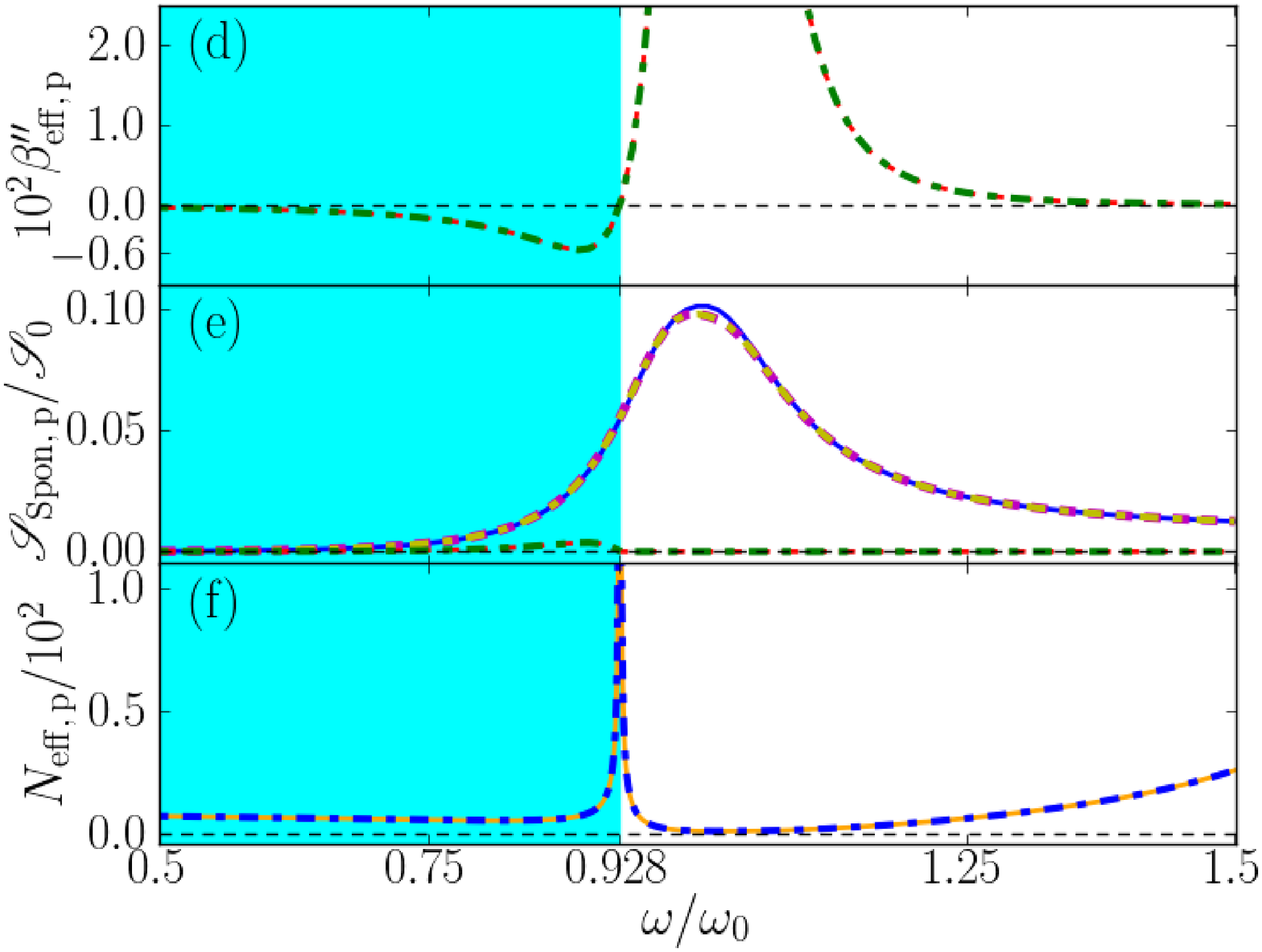}
\caption{
Power spectrum~(\ref{Power spectrum of spontaneous emission at angle theta}) of spontaneous emission of noise photons exiting a loss-compensated multilayer metamaterial at an angle of 30 degrees away from the normal, in units of ${\mathscr{S}}_0=\hbar\omega_0^3/4\pi\varepsilon_0c^3$,  due to spontaneous emission of noise photons within the loss-compensated multilayer metamaterial of Fig.~\ref{Fig:Scheme of the multilayer loss compensated medium},  at zero temperature.
Left and right panels correspond to $s$- and $p$-polarized light, respectively.
The amplifying and absorbing layers are described by the Lorentz model [Eq.~(\ref{gainmodelLorentz})],  with parameters
$\omega_{\rm p_{\rm a}}/\omega_0=0.3$, $\gamma_{\rm a}/\omega_0=0.1$
for the lossy layers, and $\omega_{\rm p_{\rm b}}/\omega_0=0.25$,
$\gamma_{\rm b}/\omega_0=0.15$ for the layers with gain. We choose
$d_{\rm a,b}\omega_0/c=0.1$ and five unit cells.
The parts~(a) and (d) show the imaginary part of the normal wave-vector component $\beta_{\rm eff}$.
In panels (b) and (e), the power spectrum of the noise photons predicted with the effective-index theories is compared to the exact multilayer calculation and the QOEM theory.
For the effective-index theories, red dotted  curves are obtained by inserting effective parameters based on Eq.~(\ref{retrived index}) into Eq.~(\ref{Power spectrum of spontaneous emission at angle theta}); the green dash-dotted lines correspond to the other procedure Eq.~(\ref{Taylor expanding of Bloch index}) to obtain effective parameters.
Similarly, for QOEM theory (discussed in Sec.~\ref{Sec:effective_medium}) the magenta dashed lines are produced with Eq.~(\ref{retrived index}), and the yellow dash-dotted curves  with Eq.~(\ref{Taylor expanding of Bloch index}).
Panels (c) and (f) show the effective noise current densities $N_{\rm eff}$ of Eq.~(\ref{effective noise photon}), in  solid orange lines as obtained using the effective index of Eq.~(\ref{retrived index}) and in   dash-dotted blue curves as produced with the other procedure [Eq.~(\ref{Taylor expanding of Bloch index})] to obtain the effective index.
 }
\label{Fig:Comparing the intensity of spontaneous emission 30}
\end{figure*}
%\end{widetex}
%
%%%%%%%%%%%%%%%%%%%%%%%%%%%%%%%%%%%%%%%%%%%%%%%%%%%%%%%%%%%%%%%%%%%%%%%%
Exact multilayer theory, in particular Eqs.~(\ref{elements of Matrixc}) and~(\ref{definition of F operators}), then
 gives the following exact long expression for the
flux of noise photons emitted from the loss-compensated multilayer
\begin{widetext}
\begin{eqnarray}\label{correlation for FdagerF}
\langle F_{\sigma,+}^\dag({\bf k},\omega)\,\, F_{\sigma',+}({\bf k}',\omega')\rangle_{\rm exact} & = &2\sum_{j=2}^{N}\left\{{\rho_{\sigma,+}^{(j)}\,\sinh(\beta''_{j}l_{j}) {\rm sgn}[\varepsilon_{{\rm
I}\,j}(\omega)]\left({|{\cal B}_{\sigma\,21}^{(j)}|^2e^{-\beta''_{j}l_{j}}+|{\cal B}_{\sigma\,22}^{(j)}|^2e^{\beta''_{j}l_{j}}}\right)}\right.\nonumber\\
&&\left.{-\frac{|\beta''_{j}|}{\beta'_{j}}\,\rho_{\sigma,-}^{(j)}\, \sin(\beta'_{j}l_{j})\left({{\cal B}_{\sigma\,21}^{(j)\,*}{\cal B}_{\sigma\,22}^{(j)}e^{ {\rm i}\beta'_{j}(z_{j}+z_{j-1})}+{\cal B}_{\sigma\,21}^{(j)}{\cal B}_{\sigma\,22}^{(j)\,*}e^{- {\rm i}\beta'_{j}(z_{j}+z_{j-1})}}\right)}\right\}\nonumber\\
&&\times\left({{N_{\rm
th}(\omega,T)\Theta[\varepsilon_{{\rm I}\,j}(\omega)]}+{\big(N_{\rm
th}(\omega,|T|)+1\big)\Theta[-\varepsilon_{{\rm
I}\,j}(\omega)]}}\right)\delta_{\sigma \sigma'}\delta (\omega  - \omega ')\delta ({\bf k}-{\bf k}').\,\,\,\,\,\,
\end{eqnarray}
\end{widetext}
While this formula is valid for all temperatures, in the following we will mostly consider power spectra at zero temperature. Passive media do not emit thermal photons in that case, but amplifying layers have population inversion and their excited-state population can decay spontaneously.
In our numerical examples, we will look at the polarization-and angle-dependent power spectrum ${\mathscr{S}}_{\rm Spon,\sigma}(\theta,\omega)$ that was defined in terms of $\langle F_{\sigma,+}^\dag({\bf k},\omega)\,\, F_{\sigma',+}({\bf k}',\omega')\rangle$ in the second equality of Eq.~(\ref{Power spectrum of of spontaneous emission}), and where we assumed that only propagating modes reach the detector and thus restricted the Fourier integral to modes with $|{\bf k}|> \omega/c$.
For definiteness, for loss-compensated metamaterials at zero temperature the angle-dependent power spectrum is given by
\begin{widetext}
\begin{eqnarray}\label{Power spectrum of spontaneous emission at angle theta}
{\mathscr{S}}_{\rm Spon,\sigma}(\theta,\omega)&=& \frac{-\hbar\omega^3 \sin\theta}{2\pi\varepsilon_0c^3\, }\sum_{j=1}^{\frac{N}{2}-1}\left\{{\varrho_{\sigma,+}^{(2j+1)}\,\sinh(\beta''_{2j+1}d_{\rm g}) \left({|{\cal B}_{\sigma\,21}^{(2j+1)}|^2e^{-\beta''_{2j+1}d_{\rm g}}+|{\cal B}_{\sigma\,22}^{(2j+1)}|^2e^{\beta''_{2j+1}d_{\rm g}}}\right)}\right.\nonumber\\
&&\hspace{2.5cm}+\varrho_{\sigma,-}^{(2j+1)}\,\frac{|\beta''_{2j+1}|}{\beta'_{2j+1}}\, \sin(\beta'_{2j+1}d_{\rm g})\left({{\cal B}_{\sigma\,21}^{(2j+1)\,*}{\cal B}_{\sigma\,22}^{(2j+1)}e^{ {\rm i}\beta'_{2j+1}(z_{2j+1}+z_{2j})}}\right.\nonumber\\
&&\hspace{6.5cm}+\left.{\left.{{\cal B}_{\sigma\,21}^{(2j+1)}{\cal B}_{\sigma\,22}^{(2j+1)\,*}e^{- {\rm i}\beta'_{2j+1}(z_{2j+1}+z_{2j})}}\right)}\right\}.
\end{eqnarray}
\end{widetext}
This formula predicts the output intensity of light by summing up all processes by which  photons are spontaneously generated in the amplifying layers and, after reflections and transmissions, with some probability end up in the output channel of our interest.

What corresponding power spectrum does the quantum optical effective-index theory of Sec.~\ref{Sec:effective_index} predict?
From the definitions~(\ref{Neffsp}) together with the commutation relation~(\ref{commutation_Feff}), the flux of noise photons emitted by the multilayer slab at a finite temperature $T$ can within the effective-index theory be expressed in terms of the effective reflection and transmission amplitudes as
\begin{eqnarray}\label{eq:F eff}
&&\langle F^\dag_{{\rm eff\, \sigma},\pm}({\bf k},\omega) \,F_{{\rm eff\,\sigma'}, \pm}({\bf k'},\omega')\rangle_{\rm QOEI}=
\nonumber \\
&& \left\{{{N_{\rm
th}(\omega,T)\Theta[\varepsilon_{{\rm I\,eff}}(\omega)]}-{\big(N_{\rm
th}(\omega,|T|)+1\big)\Theta[-\varepsilon_{{\rm
I\,eff}}(\omega)]}}\right\} \nonumber\\
&&\times(1-|r_{\rm eff,\,\sigma}|^2-|t_{\rm eff,\,\sigma}|^2) \delta_{\sigma\,\sigma'}\delta({\bf k}-{\bf k'})\delta(\omega-\omega').
\end{eqnarray}
Here  $k_{\rm B}$  is the Boltzmann constant and $T$ is temperature, and $N_{\rm th}=1/(\exp[\hbar\omega/k_B T]-1)$ is the thermal distribution of photon  states at energy $\hbar \omega$. Notice that this flux of noise photons in Eq.~(\ref{eq:F eff}) is always a non-negative quantity (as it should be): for media that are effectively absorbing at frequency $\omega$, the $\varepsilon_{{\rm I\,eff}}(\omega)$ is positive and so is $(1-|r_{\rm eff,\,\sigma}|^2-|t_{\rm eff,\,\sigma}|^2)$, while for effectively amplifying media, both these quantities are negative.
The form of the power spectrum of the spontaneously emitted light~(\ref{Power spectrum of of spontaneous emission}) within the effective-index theory is now obtained by substituting Eqs.~(\ref{eq:F eff}) and~(\ref{Eq:classical_r_and_t}) into Eq.~(\ref{Power spectrum of of spontaneous emission}).
%
%%%%%%%%%%%%%%%%%%%%%%%%%%%%%%%%%%%%%%%%%%%%%%%%%%%%%%%%%%%%%%%%%%%%%%%%
\begin{figure*}[t]
%\begin{minipage}[b]{0.495\linewidth}
%\begin{minipage}[b]{1.0\linewidth}
%\centering
%\includegraphics[width=\textwidth]{IS60.eps}
%\includegraphics[width=\textwidth]{IS60.png}
\includegraphics[width = \columnwidth]{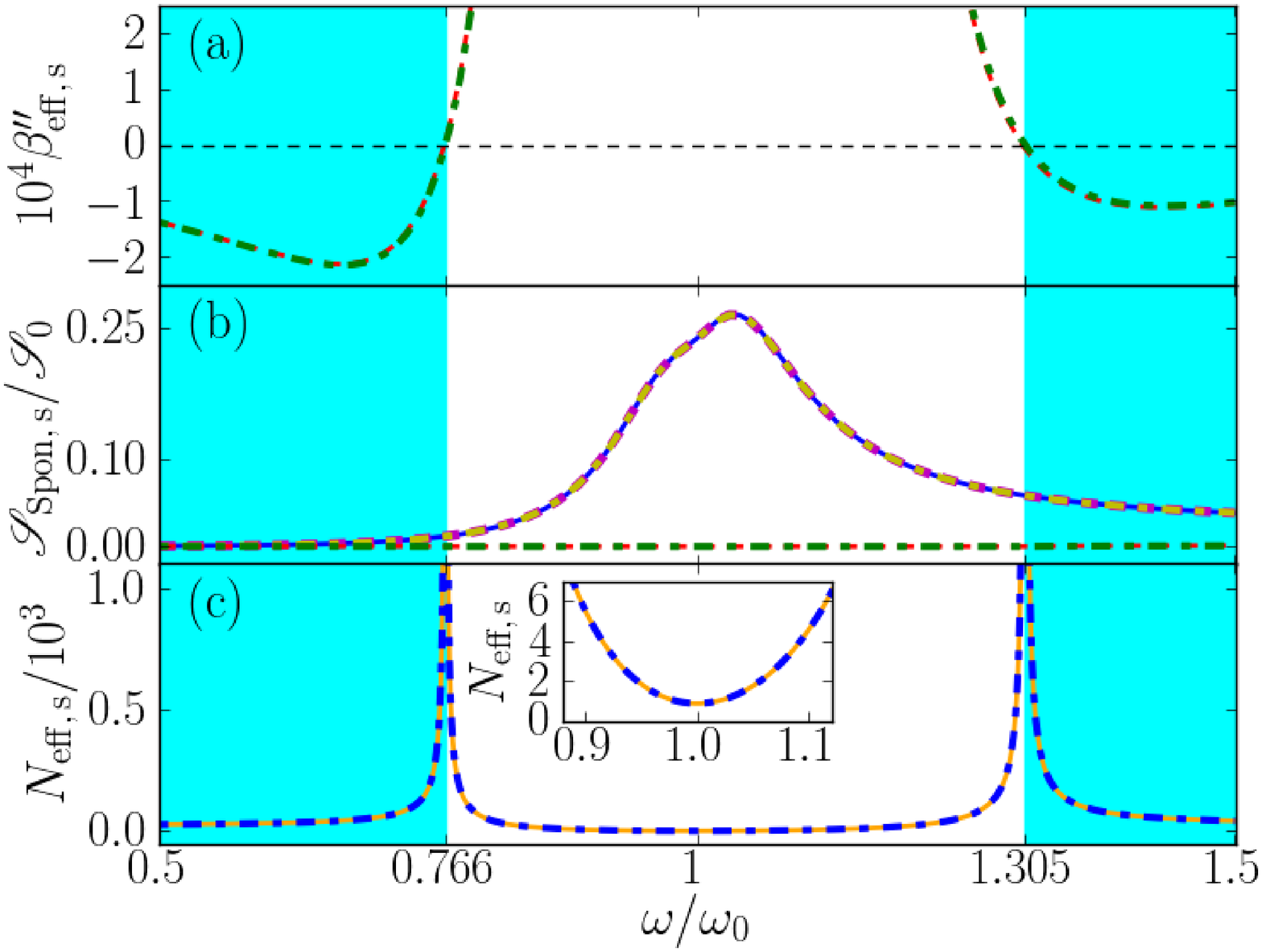}
%\end{minipage}
%\hspace{0cm}
%\begin{minipage}[b]{0.49\linewidth}
%\begin{minipage}[b]{1.0\linewidth}
%\centering
%\includegraphics[width=\textwidth]{IP60.eps}
%\includegraphics[width=\textwidth]{IP60.png}
\includegraphics[width = \columnwidth]{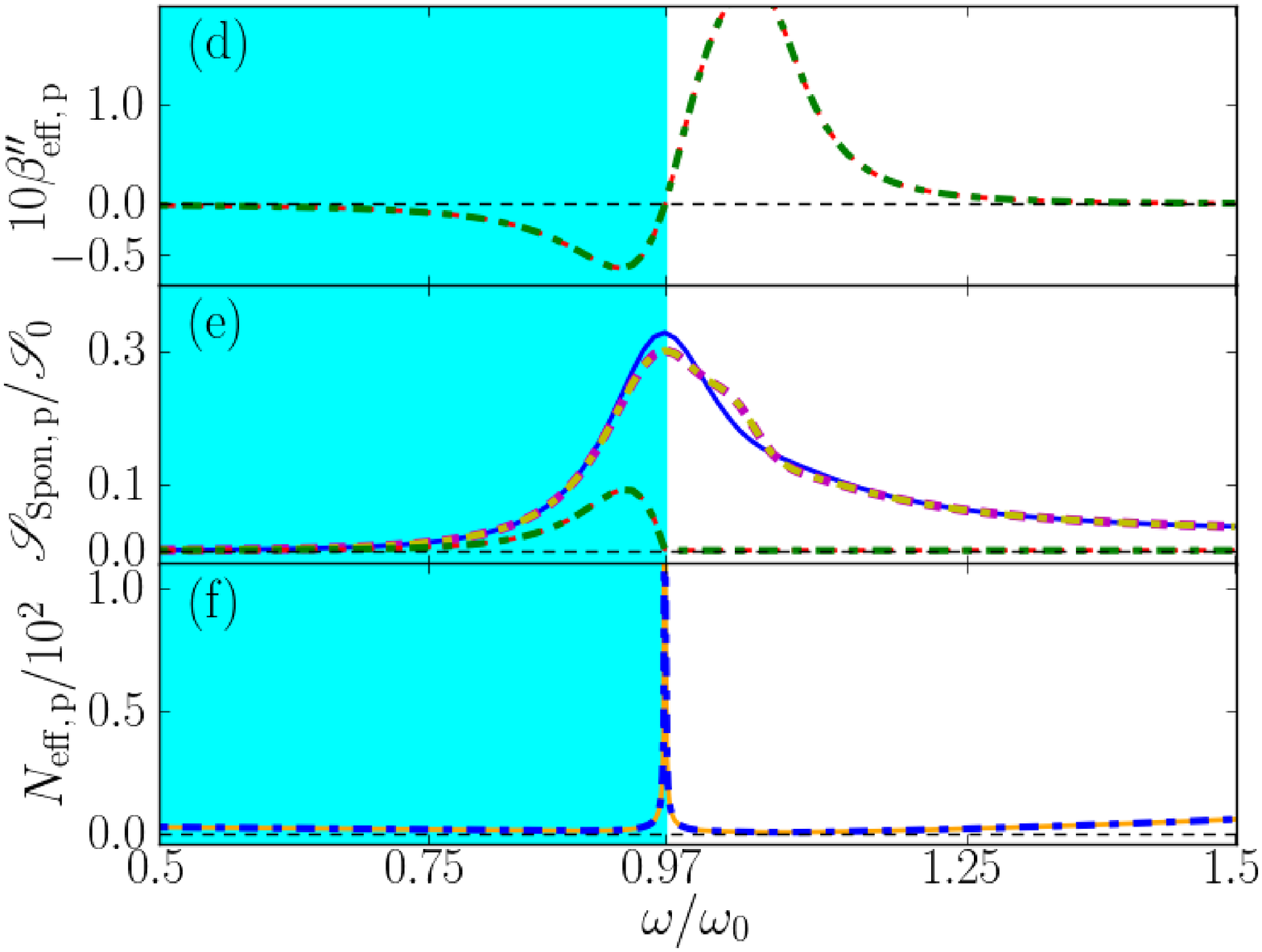}
%\end{minipage}
\caption{As in Fig.~(\ref{Fig:Comparing the intensity of spontaneous emission 30}) but now for light emission  at an angle of 60 degrees with respect to the normal. }
\label{Fig:Comparing the intensity of spontaneous emission 60}
\end{figure*}
%
%%%%%%%%%%%%%%%%%%%%%%%%%%%%%%%%%%%%%%%%%%%%%%%%%%%%%%%%%%%%%%%%%%%%%%%%

To be specific, unless stated explicitly, below in our numerical examples we will assume the temperature to be zero Kelvin.
Furthermore, we will assume that lossy and amplifying layers can be described by Lorentzian dielectric functions. A medium consisting of two-level atoms with a population $N_{\rm up}$ in the upper level and $N_{\rm down}$ in the lower level can near its resonance frequency $\omega_0$ be described by an electric permittivity  of the Lorentzian form~\cite{Matloob1997a}
\begin{eqnarray}\label{gainmodelLorentz}
\varepsilon(\omega)  =  1+\left({\frac{N_{\rm down}-N_{\rm up}}{N_{\rm down}+N_{\rm up}}}\right)\frac{\omega_{\rm p}^2}{\omega_{0}^2-\omega^2-\imath\gamma\omega}.
\end{eqnarray}
where $\omega_{\rm p}$ is the
coupling frequency, $\omega_{0}$ is the transverse resonance
frequency, and $\gamma$ is the dissipation
and amplification parameters for lossy and amplifying layers, respectively.
The population factor that occurs in the dielectric function~(\ref{gainmodelLorentz}) is positive for passive systems, $N_{\rm down}>N_{\rm up}$, but  negative for optical gain that arises from population inversion in the medium, $N_{\rm up}>N_{\rm down}$. In addition, this factor can be written in term of the thermal distribution $N_{\rm th}$ as $(2N_{\rm th}(\omega,T)+1)^{-1}$ for lossy and as $(-2N_{\rm th}(\omega,|T|)-1)^{-1}$ for amplifying layers.
In Figure~\ref{Fig:Comparing the intensity of spontaneous emission 30}
we explore regions with net loss and net gain and the frequencies of exact loss compensation that separate them, and study the corresponding flux of noise photons and the effective noise photon distribution, all corresponding to an output angle of $30^\circ$. Left panels depict $s$- and right panels $p$-polarization. In  Figure~\ref{Fig:Comparing the intensity of spontaneous emission 60}

we show the analogous results for an emission angle of $60^\circ$. (Some panels of the two figures will be explained below in Sec.~\ref{Sec:effective_medium}.)

Exact loss compensation occurs when the imaginary part of the normal-wave vector components $\beta_{\rm eff}$ vanishes. We show $\beta_{\rm eff}$  in panels~(a) and (d) of both figures, where it is also clear that the two methods to retrieve effective parameters lead to nearly identical results. For $s$-polarization, it follows from Eq.~(\ref{Taylor expanding of Bloch index}) that exact loss compensation occurs at angle-independent frequencies. A comparison of the panels~(a) of Figs.~\ref{Fig:Comparing the intensity of spontaneous emission 30} and \ref{Fig:Comparing the intensity of spontaneous emission 60} illustrates this, where
for the parameters chosen, exact loss compensation occurs  at $0.766358\omega_0$ and $1.30487\omega_0$, net loss  in the frequency range $0.766358<\omega/\omega_0<1.30487$ and net gain at elsewhere.

By contrast, for $p$-polarized light exact loss compensation does depend on the angle of incidence, as again follows from Eq.~(\ref{Taylor expanding of Bloch index}) and as illustrated in Figs.~\ref{Fig:Comparing the intensity of spontaneous emission 30} and ~\ref{Fig:Comparing the intensity of spontaneous emission 60}: in Fig.~\ref{Fig:Comparing the intensity of spontaneous emission 30}(d) exact loss compensation occurs (only) at $\omega=0.928424\omega_0$, with net gain at smaller and net loss at higher frequencies. At sixty degrees, Fig.~\ref{Fig:Comparing the intensity of spontaneous emission 60}(d) shows exact loss compensation at a slightly higher frequency.

In panels (b) for $s$-polarization and (e) for $p$-polarization, power spectra are displayed for spontaneously emitted light that exits the metamaterial at an angle of $30^\circ$ (Fig.~\ref{Fig:Comparing the intensity of spontaneous emission 30}) and $60^\circ$ (Fig.~\ref{Fig:Comparing the intensity of spontaneous emission 60}). Note that these angular power spectra are continuous also across  frequencies of exact loss compensation.

It is known that for lossy homogeneous media at zero temperature the flux of noise photons vanishes, so the power spectrum of the outgoing noise photons vanishes. The effective-index theory predicts something else, namely that no photons are emitted by {\em effectively} lossy loss-compensated metamaterials. This prediction is illustrated in panels~(b) and (e) of Figs.~\ref{Fig:Comparing the intensity of spontaneous emission 30} and \ref{Fig:Comparing the intensity of spontaneous emission 60}, especially around $\omega_0$ for outgoing $s$-polarized light, and above $0.928424\omega_0$ for outgoing $p$-polarized light. By contrast, the full gain-loss multilayer calculation does predict the emission of noise photons at zero temperature, as the figures show. Thus effective-index theory clearly fails for effectively lossy loss-compensated metamaterials. At exact loss compensation ($\varepsilon_{{\rm I\,eff}}=0$), by  Eq.~(\ref{eq:F eff}) the effective-index theory  predicts that the flux of noise photons vanishes, which the figures show is another failure of the effective-index theory.

For effectively amplifying loss-compensated  metamaterials, the effective-index theory does predict a finite flux of spontaneously emitted photons that grows with the effective gain, as is best visible in Fig.~\ref{Fig:Comparing the intensity of spontaneous emission 60}(d), where for frequencies slightly below $0.97 \omega_{0}$ much loss is  slightly overcompensated by much gain. Again the effective-index theory is clearly far from accurate.
Hence, for loss-compensated metamaterials at zero temperature, we find a clear failure of the quantum optical effective-index theory to predict an accurate power spectrum for loss-compensated metamaterials. A new effective theory is needed that also accurately describes the amount of noise photons in metamaterials.

%%%%%%%%%%%%%%%%%%%%%%%%%%%%%%%%%%%%%%%%%%%%%%%%%%%%%%%%%%%%%%%%%%%%%
\section{Quantum optical effective-medium theory}\label{Sec:effective_medium}
%%%%%%%%%%%%%%%%%%%%%%%%%%%%%%%%%%%%%%%%%%%%%%%%%%%%%%%%%%%%%%%%%%%%%
We will now derive a quantum-optical effective medium (QOEM)
theory that does give accurate predictions for thee-dimensional light propagation in loss-compensated
media. In contrast to the previous effective theory, it is not an effective-index theory, because besides the effective index another effective parameter will be needed. Our approach is
to distill solely from a unit cell not only the usual $\beta_{\rm eff,\, \sigma }$
but also an effective noise photon distribution $N_{\rm eff,\, \sigma}({\bf k},\omega,T)$. The theory presented here is a generalization of Ref.~\onlinecite{Amooghorban2013a}, which is valid for normal incidence, to arbitrary angles of incidence.

Analogous to effective-index theory, we will again assume that there is an effective noise operator in the unit cell. However, unlike in the effective-index theory we will not try not define this operator, but rather determine the expectation value of its corresponding number operator. Analogous to Eq.~(\ref{eq:F eff}) for the effective-index theory, we will write the expectation value
\begin{widetext}
\begin{eqnarray}\label{definition noise for unit cell}
\langle F^\dag_{\rm \sigma}({\bf k},\omega)F_{\rm \sigma'}({\bf k'},\omega')\rangle_{\rm QOEM}
&=&\left\{{{N_{\rm eff,\, \sigma}({\bf k},\omega,T)\Theta[\varepsilon_{{\rm unit,eff\,
I}}(\omega)]}-{\big(N_{\rm eff,\, \sigma}({\bf k},\omega,|T|)+1\big) \Theta[-\varepsilon_{{\rm unit,eff\,
I}}(\omega)]}}\right\} \nonumber\\
& \times &\left(1-|R_{{\rm unit,eff
},\,\sigma}|^2-|T_{{\rm unit,eff
},\,\sigma}|^2\right) \delta_{\sigma\,\sigma'}\delta({\bf k}-{\bf k'})\delta(\omega-\omega'),
\end{eqnarray}
\end{widetext}
in terms of the effective noise current density $N_{\rm eff}$ that we define shortly. The  $R_{{\rm unit,eff},\,\sigma}$ and $T_{{\rm unit,eff},\,\sigma}$ are the (classical) reflection and transmission amplitudes of the entire unit cell.  If the factor  $(1-|R_{{\rm unit,eff},\,\sigma}|^2-|T_{{\rm unit,eff},\,\sigma}|^2)$ is positive then  it quantifies the amount of net absorption in the unit cell, otherwise it quantifies the net amount of amplification.  The difference with Eq.~(\ref{eq:F eff}) for the effective-index theory that featured thermal distributions $N_{\rm th}$ is thus that here instead we defined an effective  distribution that in general is not a thermal one.

We fix $\langle F^\dag F\rangle_{\rm QOEM}$ of Eq.~(\ref{definition noise for unit cell}) and thereby $N_{\rm eff}$ in three steps. First, we apply our general input-output theory of Sec.~\ref{Sec:Multilayer} to a single unit cell of the metamaterial. Second, we require that the expectation value $\langle F^\dag F\rangle_{\rm QOEM}$ coincides with the corresponding unit-cell-averaged noise expectation value of the general multilayer theory.
Third, we  make use of our assumption that the unit cell of the metamaterial is much thinner than an optical wavelength, so we can Taylor expand the results from multilayer theory to first order in the layer thicknesses $d_{\rm a,b}$ and obtain
\begin{eqnarray}\label{noise contribution for unit cell}
&&\langle F^\dag_{\rm \sigma}({\bf k},\omega)F_{\rm \sigma'}({\bf k'},\omega')\rangle_{\rm QOEM} =  \sum_{j={\rm a,b}} \frac{d_{j}\,|\varepsilon_{j,I}|\,\omega^2\,{\cal K}_{j,\sigma}(\theta)}{c^2\beta_0}\,\nonumber\\
&& \times N_{\rm th}(\omega,|T_j|)\,\delta_{\sigma\,\sigma'}\delta({\bf k}-{\bf k'})\delta(\omega-\omega'),
\end{eqnarray}
where ${\cal K}_{j,s}(\theta)=1$ and ${\cal K}_{j,p}(\theta)= ({\sin^2 \theta}+|\varepsilon_j|^2{\cos^2 \theta})/|\varepsilon_j|^2$. Now we have two expressions for
$\langle F^\dag F\rangle_{\rm QOEM}$, namely Eqs.~(\ref{definition noise for unit cell}) and (\ref{noise contribution for unit cell}). By equating these two, Taylor approximating also the net gain or loss factor $\left(1-|R_{{\rm unit,eff},\,\sigma}|^2-|T_{{\rm unit,eff
},\,\sigma}|^2\right)$ of Eq.~(\ref{definition noise for unit cell})  to first order in the unit cell thickness $d = d_{\rm a} + d_{\rm b}$,   and solving for $N_{\rm eff,\, \sigma}({\bf k},\omega)$, we obtain as a main result this effective noise photon distribution
\begin{eqnarray}\label{effective noise photon}
N_{\rm eff,\, \sigma}%({\bf k},\omega,T)
=\left\{ \begin{array}{c}
%{*{20}c}
   {\sum_{j={\rm a,b}}\eta_{j,\sigma}[N_{\rm th}(\omega,T_j)]} %&& {\rm for\,\, loss-loss\, \,metamaterials}
   \\
   {-1+\sum_{j={\rm a,b}}\eta_{j,\sigma}[N_{\rm th}(\omega,|T_j|)+1]}
   %&& {\rm for\,\, gain-gain\, \,metamaterials}
   \\
 {-\frac{1}{2}+\frac{1}{2}\sum_{j={\rm l,g}}\eta_{j,\sigma}[2N_{\rm th}(\omega,|T_j|)+1]}
  % && {\rm for\,\, loss-compensated\, \,metamaterials}
\end{array} \right.
\end{eqnarray}
which correspond, from top to bottom, to loss-loss, gain-gain, and loss-compensated metamaterials.
This effective noise photon density thus depends on the same variables as the classical effective parameter $\beta_{\rm eff}$: on the angle of incidence, on the polarization of the input state, as well as on the dielectric parameters of the unit cell via
\begin{eqnarray}\label{parameter eta}
\eta_{j,\sigma}(\theta)=p_j \,\frac{{\cal K}_{j,\sigma}(\theta)}{{\cal K}_{\rm eff,\sigma}(\theta)}\left|{\frac{\varepsilon_{j,\,{\rm I}}(\omega)}{\varepsilon_{{\rm eff\, ,I}}(\omega)}}\right|,
\end{eqnarray}
where the $p_{j} = d_{j}/d$ are the volume fractions of the layers and ${\cal K}_{\rm eff,\sigma}(\theta)$ equals ${\cal K}_{j,\sigma}(\theta)$ with $\varepsilon_j$ replaced by $\varepsilon_{\rm eff}$. We allowed the two types of layers of the unit cell to be at different temperatures. Generalizations to more than two layers are straighforward.

Let us first apply this QOEM theory to loss-compensated metamaterials.
In panels (b) and (e) of Figs.~\ref{Fig:Comparing the intensity of spontaneous emission 30} and \ref{Fig:Comparing the intensity of spontaneous emission 60} we showed power spectra, and found that the exact multilayer theory predicts quite a lot more noise photons than the effective-index theory did. By contrast, the curves of QOEM theory almost coincide with the exact multilayer result in all these panels. This illustrates that we have an accurate quantum optical effective-medium theory for three-dimensional light propagation in multilayer metamaterials, both for $s$- and $p$-polarized light.
%%%%%%%%%%%%%%%%%%%%%%%%%%%%%%%%%%%%%%%%%%%%
%
\begin{figure*}[t]
%\includegraphics[width=1.0\columnwidth]{Eta.eps}
%\includegraphics[width=0.8\columnwidth]{Eta.png}
%\begin{minipage}[b]{0.495\linewidth}
%\begin{minipage}[b]{0.99\linewidth}
%\centering
%\includegraphics[width=\textwidth]{Eta.eps}
%\includegraphics[width=\textwidth]{Eta.png}
\includegraphics[width = \columnwidth]{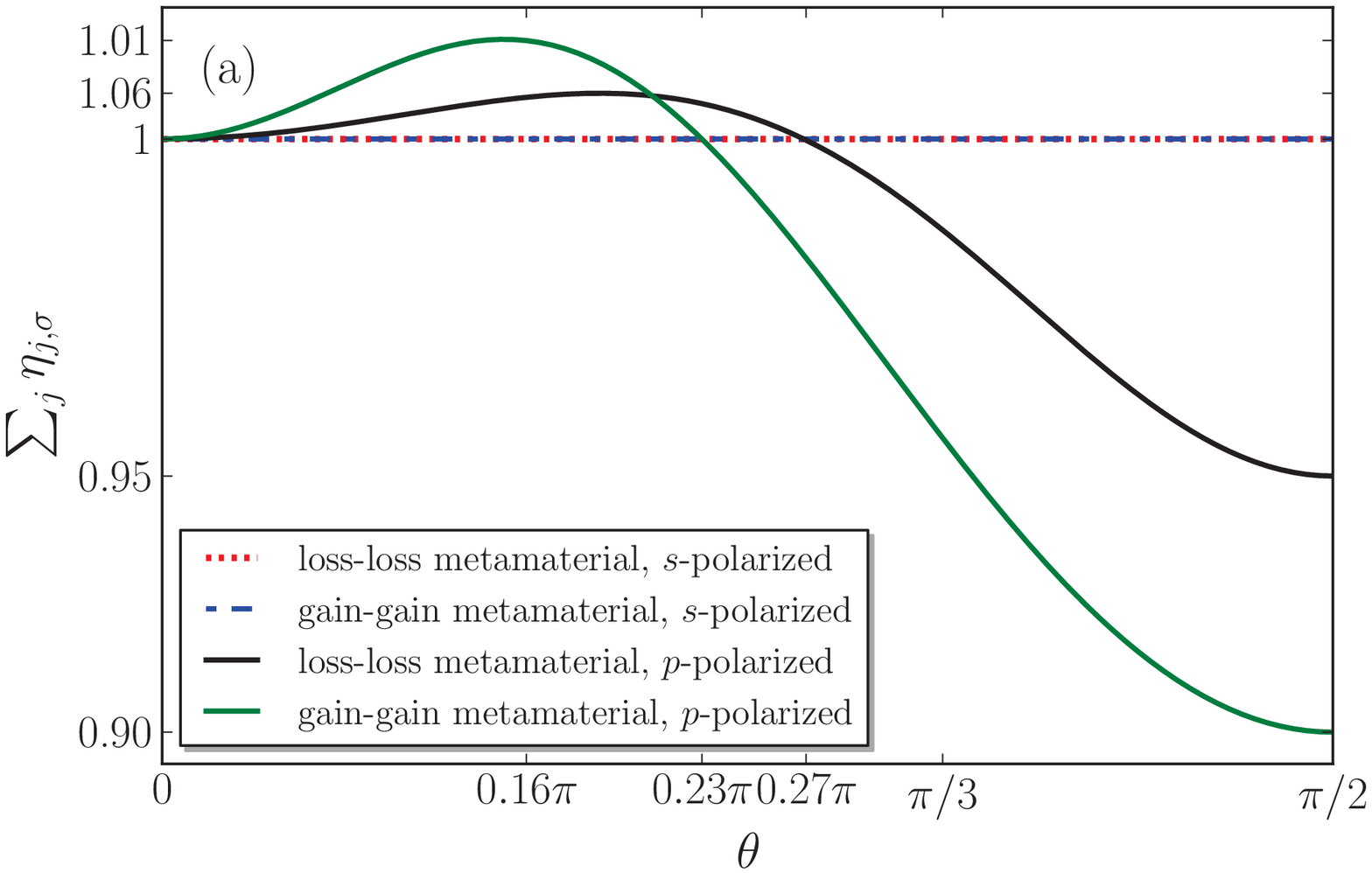}
%\end{minipage}
%\hspace{0cm}
%\begin{minipage}[b]{0.49\linewidth}
%\begin{minipage}[b]{0.99\linewidth}
%\centering
%\includegraphics[width=\textwidth]{Eta3Dgain.eps}
%\includegraphics[width=\textwidth]{Eta3Dgain.png}
%\end{minipage}
\includegraphics[width = \columnwidth]{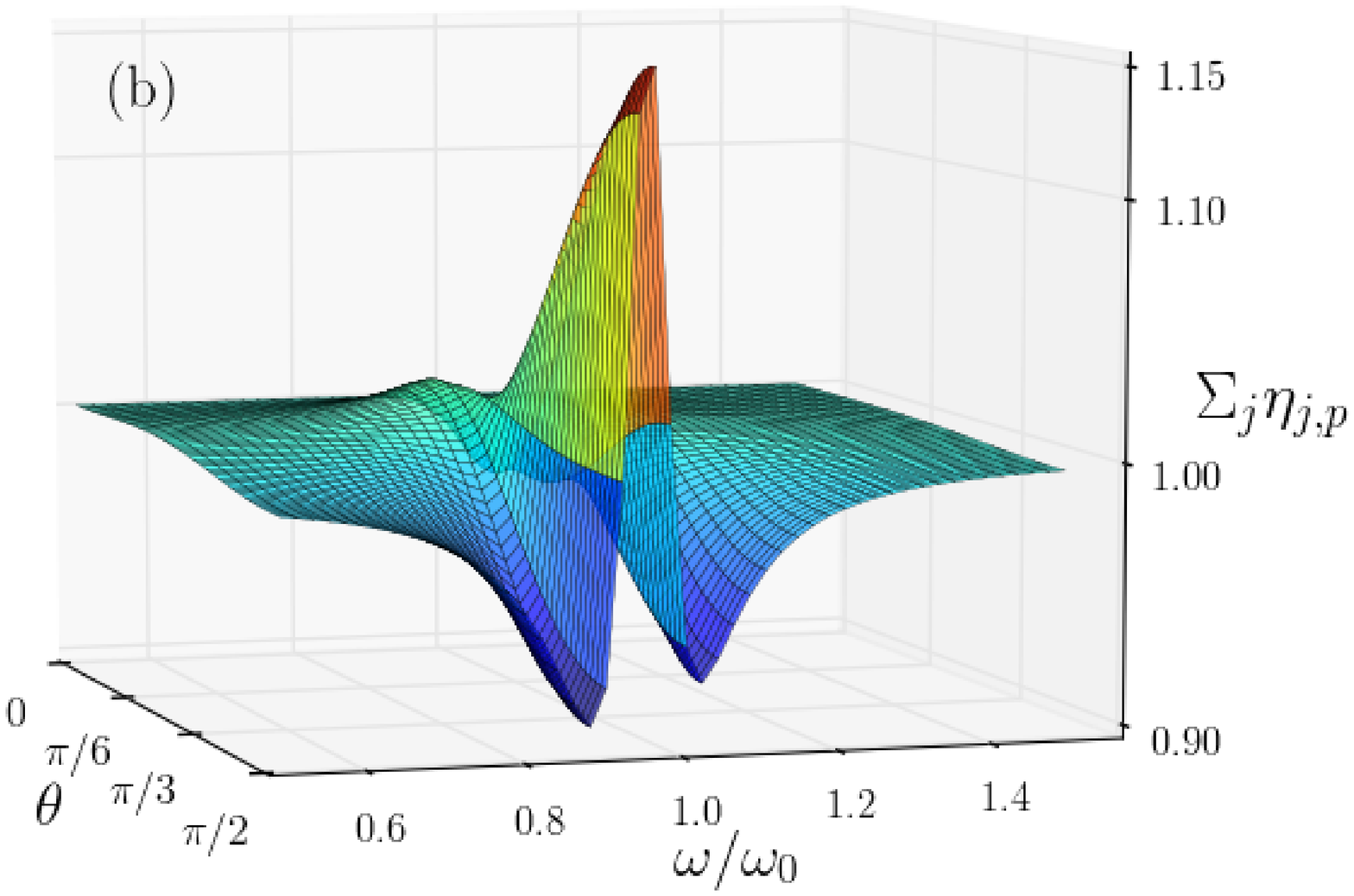}
\caption{(Color online) (a) The sum $\Sigma_{j=a,b}
\eta_{j,\sigma} $ is shown as a function of the angle of incidence $\theta$ for $s$-polarized component of input light impinging on the loss-loss (dotted red line) and the gain-gain (dash-dotted blue line) multilayers, and $p$-polarized component of input light incident on the loss-loss (solid black line) and the gain-gain (solid green line) multilayers. (b) The sum $\Sigma_{j=a,b}
\eta_{j,p} $ is shown as a function of the angle of incidence $\theta$ and of the dimensionless frequency $\omega/\omega_0$ for $p$-polarized component of input light impinging on a gain-gain multilayer.
The multilayer metamaterial with geometry of Fig.~\ref{Fig:Scheme of the multilayer loss compensated medium} has alternating layers with equal thickness  $d_{\rm a,b}\omega_0/c=0.1$, with dielectric parameters in Eq.~(\ref{gainmodelLorentz}): $\omega_{\rm p_{\rm a}}/\omega_0=0.3$, $\omega_{\rm p_{\rm b}}/\omega_0=0.1$ and $\gamma_{\rm a,b}/\omega_0=0.1$. In panel (a), we choose $\omega/\omega_0=0.9$.}
\label{Fig:Fig:Sum of Parameter Eta}
\end{figure*}
%
%%%%%%%%%%%%%%%%%%%%%%%%%%%%%%%%%%%%%%%%%%%%

What do we know about the new effective parameter, the effective noise photon distribution?
To gain some intuition, notice that from Eqs.~(\ref{effective noise photon}) and (\ref{parameter eta}) it follows that $N_{\rm eff}$ grows when loss in the metamaterial is more exactly compensated by gain [smaller $\varepsilon_{{\rm eff\, ,I}}(\omega)$] or when
the same value $\varepsilon_{{\rm eff\, ,I}}(\omega)$ results from compensating more loss by more gain (i.e. with  $\left|\varepsilon_{a,\,{\rm I}}(\omega)\right|)$ and $\left|\varepsilon_{a,\,{\rm I}}(\omega)\right|)$ both larger). This is illustrated in panels (c) and (f) of Figs.~\ref{Fig:Comparing the intensity of spontaneous emission 30} and \ref{Fig:Comparing the intensity of spontaneous emission 60}, where we see that $N_{\rm eff}$ even diverges at exact loss compensation, but in such a way as to keep the   power spectra at those frequencies continuous. This means that for metamaterials with more effective loss compensation, it becomes increasingly important to use $N_{\rm eff}$ as an additional effective-medium parameter instead of $N_{\rm th}$. These results illustrate that we have successfully generalized the one-dimensional QOEM theory of our Ref.~\onlinecite{Amooghorban2013a} to light propagation in three dimensions.

%
% * <amoghorban@gmail.com> 2015-12-03T16:34:35.161Z:
%
% ^.
Let us now show that our new QOEM theory indeed reduces to the one of Ref.~\onlinecite{Amooghorban2013a} in case of light propagation perpendicular to the interface, i.e. for ${\bf k} = {\bm 0}$. In that case, the parameter ${\cal K}_{j,\sigma}(\theta)$ in Eq.~(\ref{noise contribution for unit cell}) tends to unity for both polarizations. This in turn implies that the parameter $\eta_{j,\sigma}(\theta)$ defined in Eq.~(\ref{parameter eta}) tends to $p_j \left|\varepsilon_{j,\,{\rm I}}(\omega)/\varepsilon_{{\rm eff\, ,I}}(\omega)\right|$. This indeed  agrees with Ref.~\onlinecite{Amooghorban2013a}, where we showed that the quantum optical effective-medium theory with $N_{\rm eff}$ for normal incidence gave accurate predictions for loss-compensated, loss-loss as well as gain-gain metamaterials.
%
%%%%%%%%%%%%%%%%%%%%%%%%%%
%%%%%%%%
\begin{figure*}[t]
\begin{minipage}[b]{0.495\linewidth}
\centering
\includegraphics[width=\textwidth]{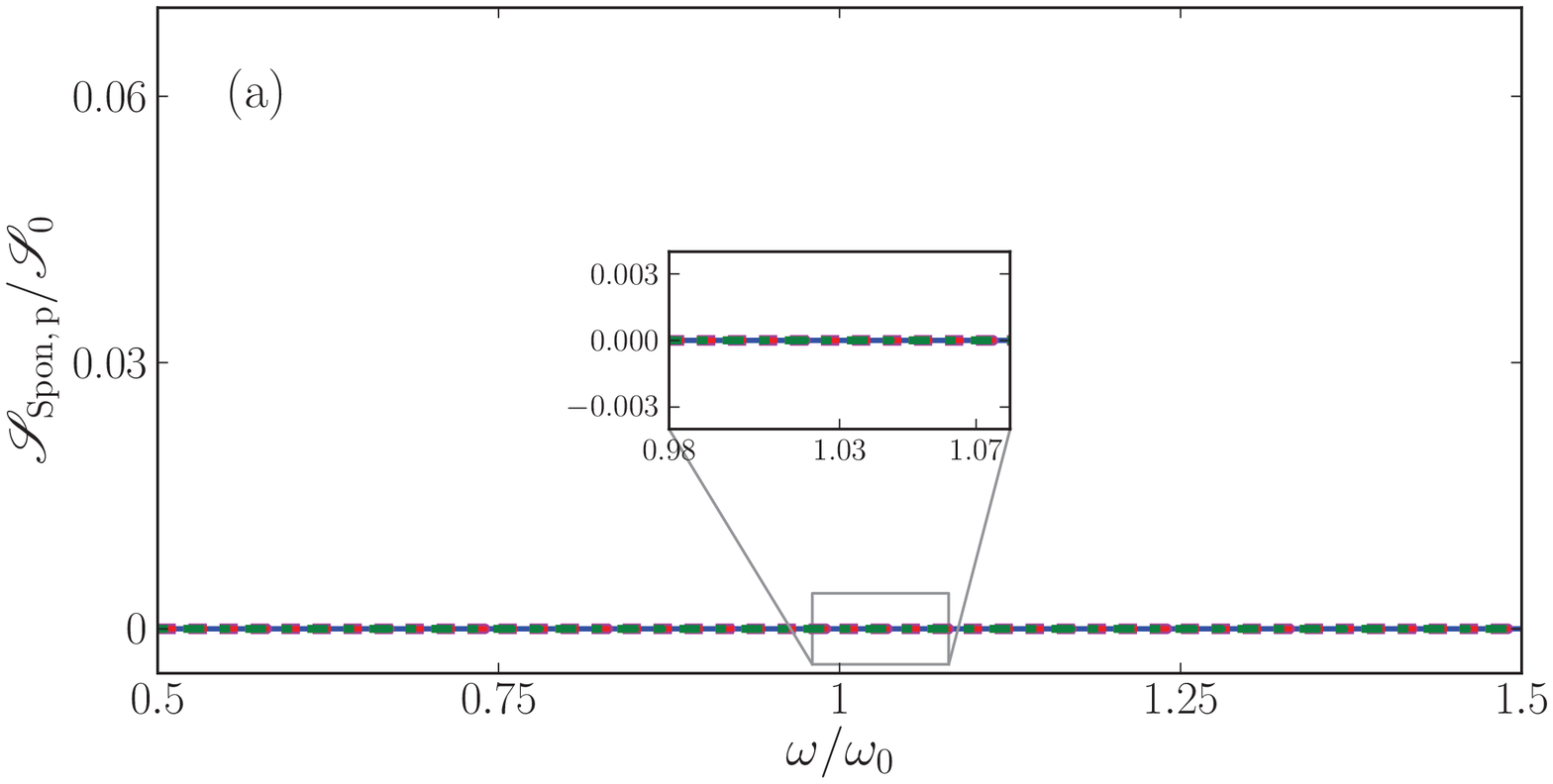}
\end{minipage}
\hspace{0cm}
\begin{minipage}[b]{0.49\linewidth}
\centering
\includegraphics[width=\textwidth]{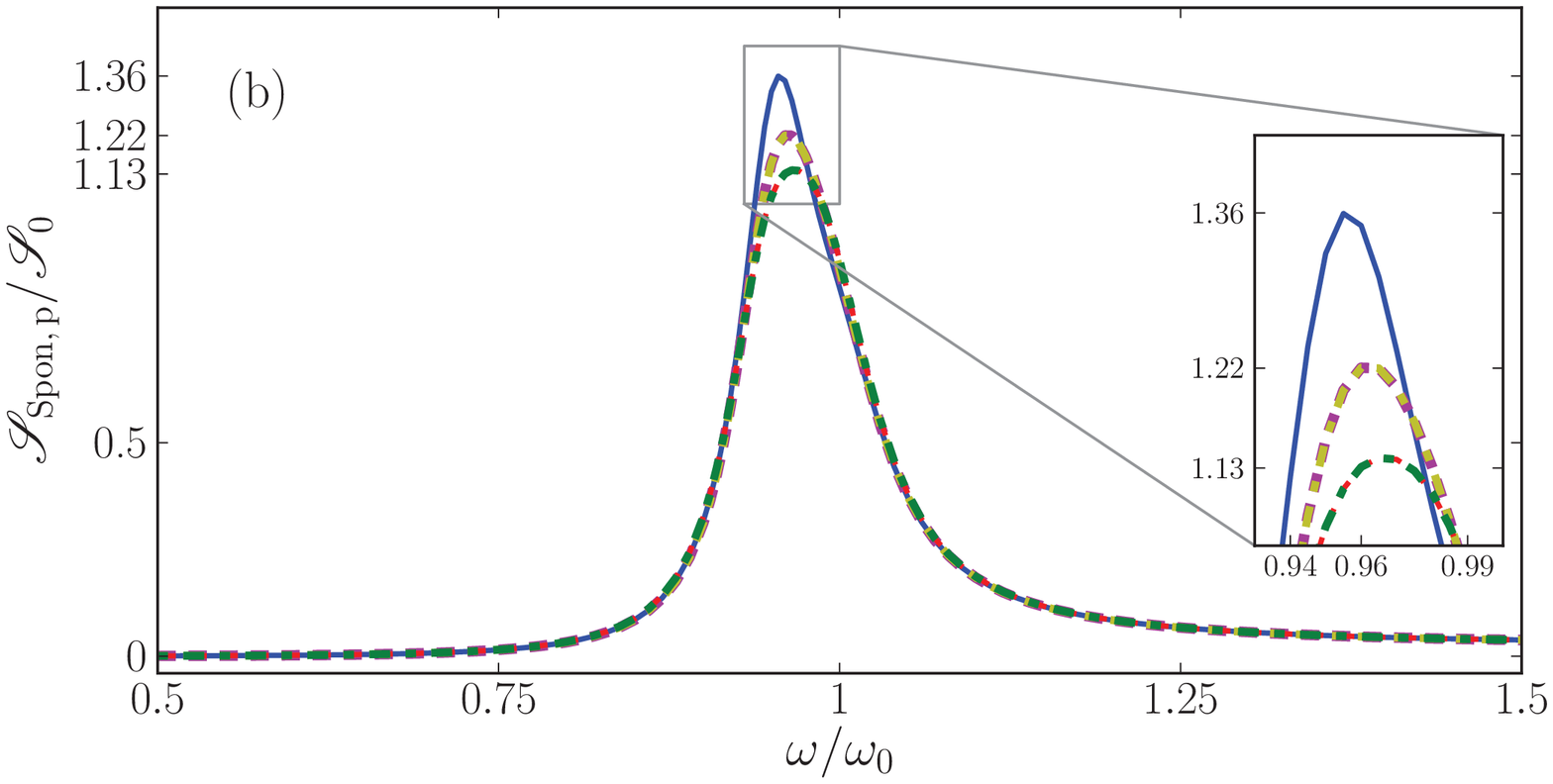}
\end{minipage}
\begin{minipage}[b]{0.495\linewidth}
\centering
\includegraphics[width=\textwidth]{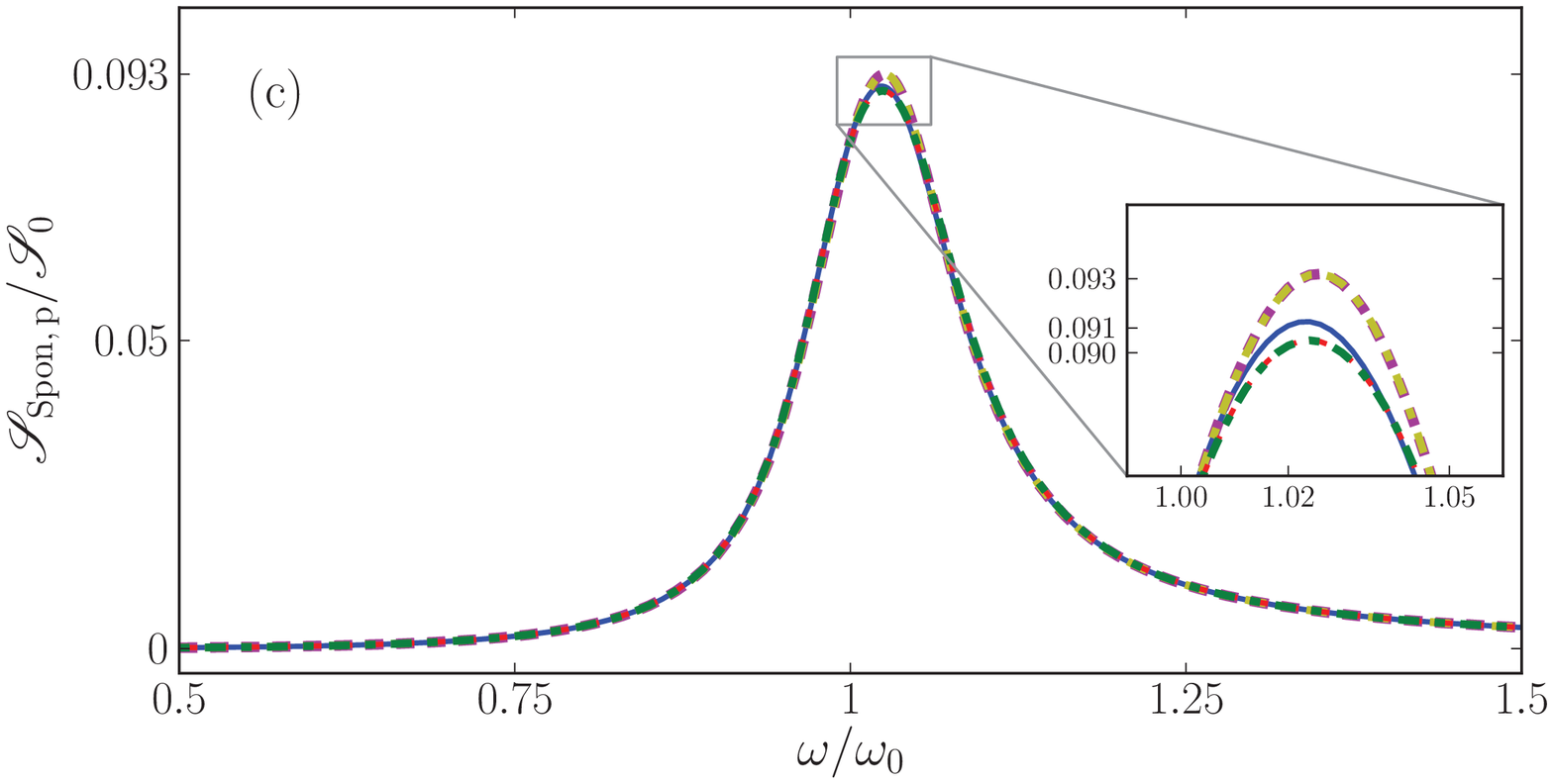}
\end{minipage}
\hspace{0cm}
\begin{minipage}[b]{0.49\linewidth}
\centering
\includegraphics[width=\textwidth]{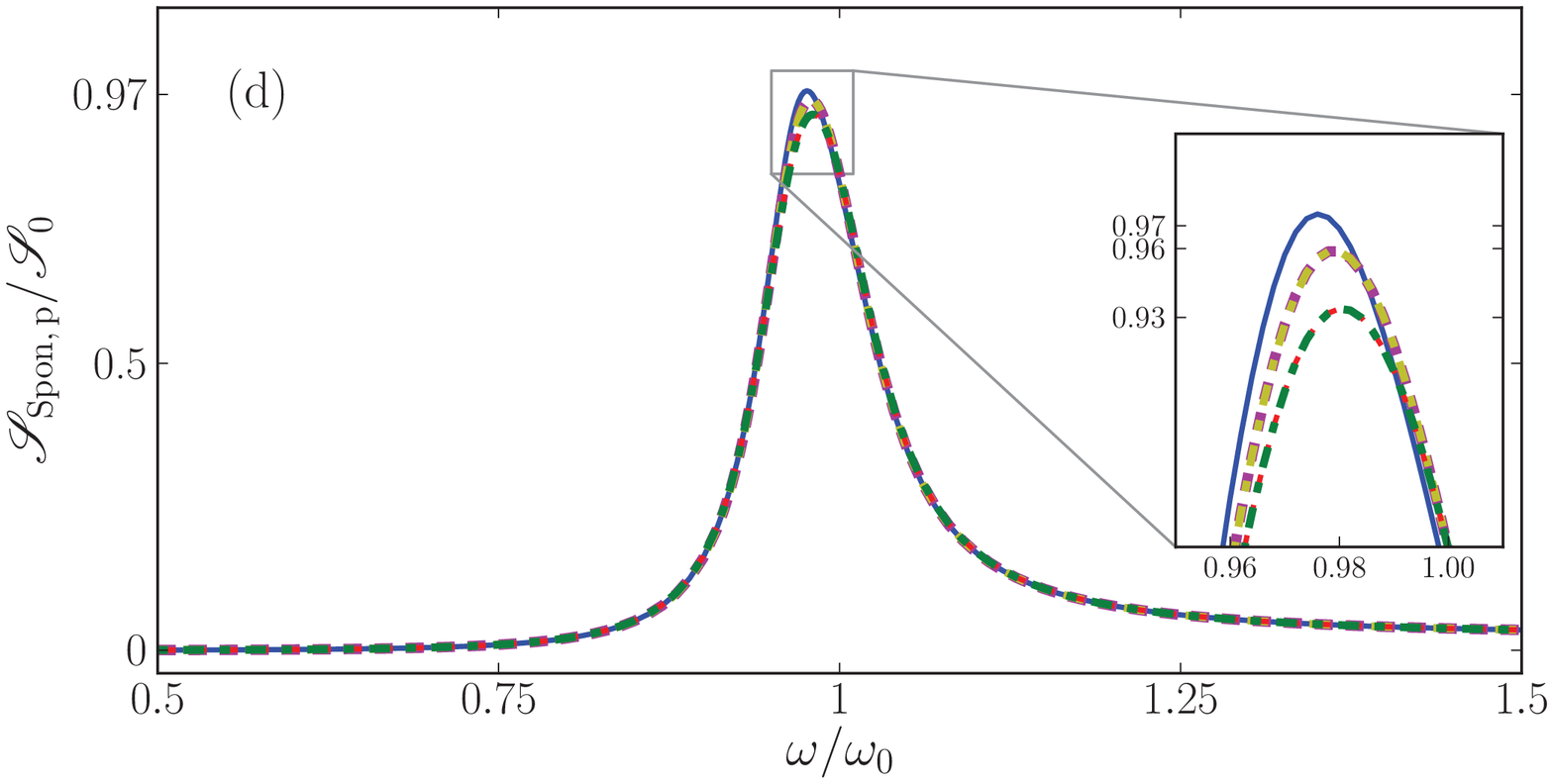}
\end{minipage}
\caption{The spontaneous-emission power spectrum of the noise photons~(\ref{Power spectrum of spontaneous emission at angle theta}), in units of ${\mathscr{S}}_0=\hbar\omega_0^3/4\pi\varepsilon_0c^3$, for $p-$ polarized  light exiting the MM at 60 degrees away from the normal direction. In all four panels, the QOEM theory and the quantum optical effective-index theory are compared to the exact multilayer calculation. For the exact multilayer calculation (solid blue curves), the loss and gain layers are described by Lorentz model with parameters are identical to those used in Fig.~\ref{Fig:Fig:Sum of Parameter Eta}.
Left and right panels correspond to loss-loss and gain-gain metamaterials with the geometry of Fig.~\ref{Fig:Scheme of the multilayer loss compensated medium}. The loss-loss and the gain-gain multilayer are maintained at zero temperature in panels~(a) and (b), and at the elevated positive temperature $T=0.6\hbar\omega_0/k_{\rm B}$  in panel~(c), and at the elevated negative temperature $|T|=0.6\hbar\omega_0/k_{\rm B}$ in~(d).
For the effective-index theories, red dotted and green dash-dotted curves present the numerical parameters as obtained from the scattering method~(\ref{retrived index}) and the dispersion method~(\ref{Taylor expanding of Bloch index}), respectively.
For QOEM theory, magenta dashed and yellow dash-dotted lines correspond to these same  classical effective parameter retrieval methods~(\ref{retrived index}) and~(\ref{Taylor expanding of Bloch index}). These effective parameters so obtained are also used to compute $N_{\rm eff}$.
}
\label{Fig:Comparing the intensity of spontaneous emission 60 for loss-loss and gain-gain}
\end{figure*}
%%%%%%%%%%%%%%%%%%%%%%%%%%%%%%%%%%%

Do we also need the QOEM theory for loss-loss or gain-gain metamaterials? If the two layers within the unit cell are somehow kept at different temperatures, then we do, but this is not easy to realize. But if the entire unit cell is kept at the same temperature, then for light propagation normal to the layers we found in Ref.~\onlinecite{Amooghorban2013a} that QOEM theory reduces to the quantum optical effective-index theory, i.e. $N_{\rm eff}$ becomes equal to the thermal noise photon distribution $N_{\rm th}$. Is this also true in three dimensions?
Let us consider $s$-polarized light first, for loss-loss and gain-gain metamaterials at a uniform temperature ($T_{\rm a}=T_{\rm b}$). This means technically that the thermal distributions in Eq.~(\ref{effective noise photon}) can be moved in front of the summation, and the remaining summation is $\sum_{j = a,b} \eta_{j,s}(\theta)$.
Since in $\eta_{j,s}(\theta)$ as defined in Eq.~(\ref{parameter eta}) the fractions ${\cal K}_{j,s}(\theta)/{\cal K}_{\rm eff,s}(\theta)$ are equal to unity, the sum $\sum_{j = a,b} \eta_{j,s}(\theta)$
becomes $\sum_{j=a,b}p_j \left|\varepsilon_{j,\,{\rm I}}(\omega)/\varepsilon_{{\rm eff\, ,I}}(\omega)\right|$, which is angle independent.
In Ref.~\onlinecite{Amooghorban2013a} we also pointed out that for normal incidence the sum $\sum_{j=a,b}p_j \left|\varepsilon_{j,\,{\rm I}}(\omega)/\varepsilon_{{\rm eff\, ,I}}(\omega)\right|$ adds up to unity for loss-loss and gain-gain metamaterials (whereas the sum is always larger than unity for loss-compensated (gain-loss) metamaterials). So now we find that for $s$-polarized light, the same relations even hold for arbitrary angles of incidence, as illustrated by the horizontal line in Figure~\ref{Fig:Fig:Sum of Parameter Eta}(a).
This angle-independence for $s$-polarized light is related to the fact that for arbitrary propagation directions the electric-field components are parallel to the interface, just like for normal incidence.
As a consequence, we find that for $s$-polarized light, the effective noise photon distribution $N_{\rm eff,s}$ reduces to the thermal distribution $N_{\rm th}$. So both for a lossy metamaterial at a homogeneous positive temperature and for an amplifying metamaterial at a spatially constant negative temperature, the QOEM theory for $s$-polarized light reduces to the quantum optical effective-index theory.

How about $p$-polarized light then, does $N_{\rm eff}$ also reduce to the thermal distribution for loss-loss and gain-gain metamaterials?
For MMs kept at a uniform temperature, the thermal factor in the summation of Eq.~(\ref{effective noise photon}) can again be put in front, so that the summation reduces to $\sum_{j = a,b} \eta_{j,p}(\theta)$. However, for $p$-polarized light the fractions ${\cal K}_{j,p}(\theta)/{\cal K}_{\rm eff,p}(\theta)$  do not give unity, and hence the sum $\sum_{j = a,b} \eta_{j,p}(\theta)$ in general does not add up to unity. The angle dependence of this sum  is illustrated in Fig.~\ref{Fig:Fig:Sum of Parameter Eta}(a). For small angles, the sum is  larger than unity while for large angles it is smaller than unity.

As a consequence, for $p$-polarized light $N_{\rm eff}$ in general does not reduce to the thermal distribution, and consequently the QOEM theory does not reduce to the quantum optical effective-index theory. For the parameters as in Fig.~\ref{Fig:Fig:Sum of Parameter Eta}, $N_{\rm eff}$ becomes larger (smaller) than the thermal distribution $N_{\rm th}$ for angles of incidence smaller (larger) than $0.27\pi$  for the loss-loss metamaterial, and for the gain-gain MM the critical angle occurs at $0.23\pi$.
This difference between the effective and thermal distributions also in metamaterials without loss compensation is a qualitatively new finding as compared to the previous results of Ref.~\onlinecite{Amooghorban2013a} for one-dimensional light propagation.

Quantitatively, the deviations of $N_{\rm eff}$ from a thermal distribution are relatively small, within ten percent for the specific dielectric parameters chosen in Fig.~\ref{Fig:Fig:Sum of Parameter Eta}(a).
The closer the sum  $\sum_j\eta_{j,p}$ comes to unity, the closer QOEM theory comes to the quantum optical effective-index theory becomes. This sum is thus a measure for the ``distance'' between the two theories, and it depends on the dielectric parameters of the unit cell. Moreover, it depends on the angle of incidence and on the frequency of light how much the two theories will agree for $p$-polarization. In order to illustrate both  dependencies, in Fig.~\ref{Fig:Fig:Sum of Parameter Eta}(b)
we depict the  sum $\Sigma_j\eta_{j,p}$ as a function of both the angle of incidence $\theta$ and of the dimensionless frequency $\omega/\omega_0$, for the same gain-gain MM as in Fig.~\ref{Fig:Fig:Sum of Parameter Eta}(a). For al combinations of parameters considered in Fig.~\ref{Fig:Fig:Sum of Parameter Eta}(b), we find that the sum has a maximal deviation from unity of around 15 percent. We find similar modest but non-negligible frequency and angle dependence (not shown) for the loss-loss multilayer of Fig.~\ref{Fig:Fig:Sum of Parameter Eta}(a).

In Figure ~\ref{Fig:Comparing the intensity of spontaneous emission 60 for loss-loss and gain-gain}
we compare power spectra for $p$-polarized light exiting at an angle of $60$ degrees away from the normal of the MM, computed with the exact multilayer theory, with QOEM theory, and with the quantum optical effective-index theory. The left panels are for loss-loss MMs, the right panels for gain-gain MMs, upper panels for zero temperature and lower panels for elevated temperature. In panel~(a) for the loss-loss multilayer at zero temperature,
quantum noise can be neglected, so just like in classical optics  the power spectrum of output light vanishes identically and perfect agreement between all curves is observed.
By contrast, for the gain-gain multilayer at zero temperature in panel~\ref{Fig:Comparing the intensity of spontaneous emission 60 for loss-loss and gain-gain}(b), the population of the two-level medium is fully inverted and the effects of quantum noise in the output cannot be neglected. The power spectrum of output light appears as a peak near $\omega_0$ which is associated with the resonance frequency of the dielectric functions of each layer.
Away from resonance, both effective theories agree with the exact calculation. Near resonance there are small  differences on the order of a few percent  between the exact multilayer calculation and the two effective-medium theories.
As seen in the zoomed inset in Fig.~\ref{Fig:Comparing the intensity of spontaneous emission 60 for loss-loss and gain-gain}(b), near the resonance the QOEM theory is slightly more accurate than the effective-index theories.

In panels~\ref{Fig:Comparing the intensity of spontaneous emission 60 for loss-loss and gain-gain}(c) for loss-loss MMs at a pretty high temperature and~\ref{Fig:Comparing the intensity of spontaneous emission 60 for loss-loss and gain-gain}(d) for gain-gain MMs at a negative temperature, the exact and the two effective power spectra again agree quite well, with only on resonance a few percent difference.  Sufficiently far from the resonance when absorption is small, the thermal noise becomes negligibly small
and the power spectrum of output noise photons is approximately zero. For the gain-gain multilayer the amplitude of the peak in panel (d) is smaller than the one in (b) since amplification within gain layers is reduced by saturation effects. We checked (but do not show it here) that these results do not depend much on the typical parameters used in Fig.~\ref{Fig:Comparing the intensity of spontaneous emission 60 for loss-loss and gain-gain}.
The overall message of Fig.~\ref{Fig:Comparing the intensity of spontaneous emission 60 for loss-loss and gain-gain} is that both the QOEM theory and the quantum optical effective-index theory are quite accurate in describing power spectra of $p$-polarized light of loss-loss and gain-gain metamaterials, with the two effective theories almost equally accurate. So one can use either $N_{\rm eff}$ or $N_{\rm th}$ as the noise photon  distribution  in Eq.~(\ref{eq:F eff}).

To summarize our findings from this section, we compared for the first time power spectra of metamaterials based on exact theory and on QOEM and effective-index theory. For loss-compensated metamaterials we find  that the effective-index theory is manifestly inadequate, both for $s$- and $p$-polarized light. By contrast, our QOEM theory in a consistent way predicts that the quantum noise contribution $\langle F^\dag_{\rm \sigma}({\bf k},\omega)F_{\rm \sigma'}({\bf k'},\omega')\rangle$ to the power spectrum of a layered metamaterial is given by Eq.~(\ref{eq:F eff}), but with the thermal distribution $N_{\rm th}(\omega,|T|)$ replaced by the effective distribution $N_{\rm eff,\, \sigma}({\bf k},\omega,T)$ of Eq.~(\ref{effective noise photon}). In the absence of loss compensation, i.e. for loss-loss and gain-gain metamaterials, we found that for $s$-polarized light the QOEM theory exactly coincides with the quantum optical effective-index theory. For $p$-polarization there is no such exact agreement in the absence of loss compensation, but numerically the differences between both effective theories are so small  that it is essentially a matter of choice which one to use. For loss-compensated metamaterials, QOEM theory is the only accurate effective-medium theory.

%%%%%%%%%%%%%%%%%%%%%%%%%%%%%%%%%%%%%%%%%%%%%%%%
\section{Second test: Propagation of squeezed states}\label{Sec:squeeze}
For the power spectra emitted by a metamaterial as discussed in Sec.~\ref{sec:intensity}, the input states of light were vacuum states, which have a classical analogue (no light). By contrast, here we analyze how well the difference effective-medium theories describe the output quantum states of light when the input states have no classical analogues.  This will serve as a useful independent test of the accuracy of the effective-medium theories.
\begin{figure}[t]
\includegraphics[width=1.0\columnwidth]{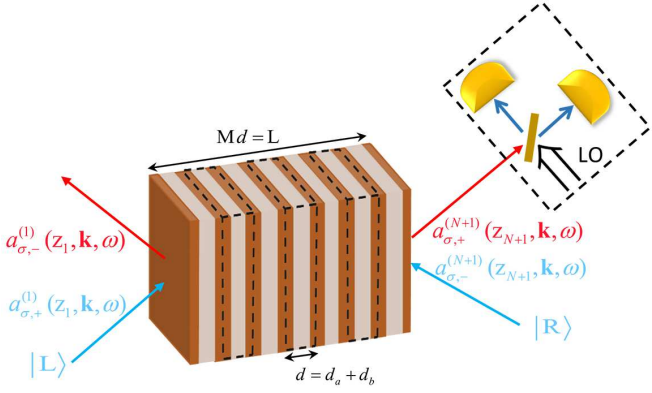}
\caption{(Color online)
Scheme of the loss-compensated multilayer  medium in air. It has alternating layers with thicknesses $d_{\rm a,b}$ that are arranged symmetrically. The two outermost layers have widths $d_{\rm a}/2$, which makes the medium finite periodic with $M$ unit cells. The amplifying and absorbing layers are described by the Lorentz model [Eq.~(\ref{gainmodelLorentz})],  with parameters
$\omega_{\rm p_{\rm a}}/\omega_0=0.3$, $\gamma_{\rm a}/\omega_0=0.1$
for the lossy layers, and $\omega_{\rm p_{\rm b}}/\omega_0=0.25$,
$\gamma_{\rm b}/\omega_0=0.15$ for the layers with gain. We choose
$d_{\rm a,b}\omega_0/c=0.1$ and $M=5$.  The incident squeezed vacuum state $| L \rangle$ has
the squeeze strength $\zeta_{\sigma}=0.2$ and phase $\phi_{\sigma,\zeta}=2\phi_{\sigma,{\rm LO}}-\frac{5}{2}$, while the squeezed vacuum state $| R \rangle$ has the same strength $\xi_{\sigma}=0.2$ with $\phi_{\sigma,\xi}=2\phi_{\sigma,{\rm LO}}-2$, all assumed to be frequency independent. The outgoing light  on the right-hand side of the multilayer metamaterial is measured with a balanced homodyne detector,  shown within the dashed box, which is assumed to  co-rotate with the angle $\theta$.
}\label{Fig:Scheme of the multilayer loss compensated medium}
\end{figure}
We will study the propagation of  squeezed states of light through the metamaterial, generalizing Ref.~\cite{Amooghorban2013a} to arbitrary angles of incidence. The main question is  how well quantum properties of the incoming state are preserved in the output, given that there is  quantum noise in the metamaterial. We compare the answers to this question as obtained by exact multilayer theory and by quantum effective-index and effective medium theories. Most importantly, we investigate whether the QOEM theory that so accurately described power spectra in Sec.~\ref{sec:intensity} also describes the propagation of squeezed states well.
\begin{figure*}[t]
%\begin{minipage}[b]{0.45\linewidth}
%\begin{minipage}[b]{0.495\linewidth}
%\centering
%\includegraphics[width=\textwidth]{DSS30.eps}
%\includegraphics[width=\textwidth]{DSS30.png}
\includegraphics[width = \columnwidth]{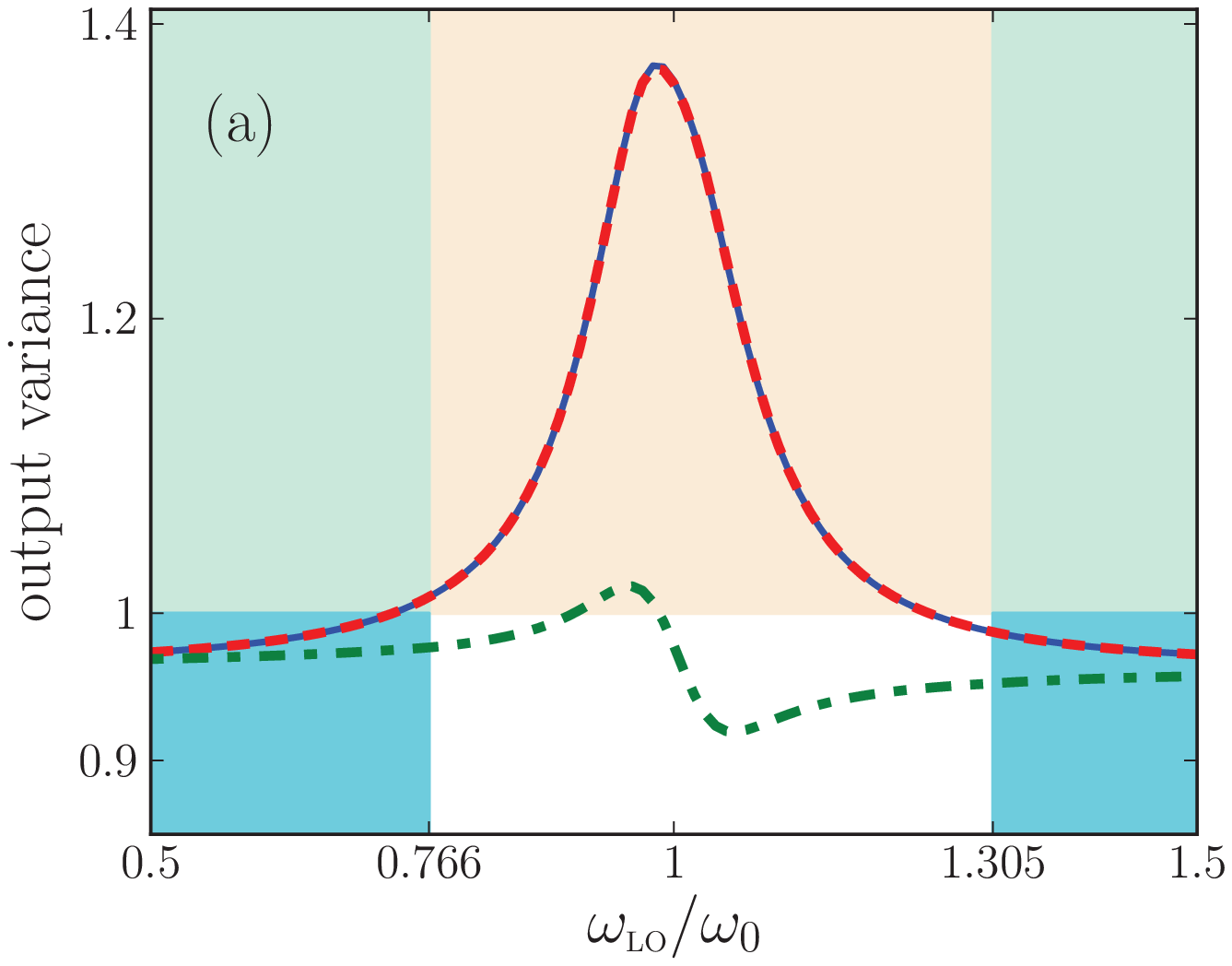}
%\end{minipage}
%\hspace{1.2cm}
%\begin{minipage}[b]{0.45\linewidth}
%\begin{minipage}[b]{0.495\linewidth}
%\centering
%\includegraphics[width=\textwidth]{DPP30.eps}
%\includegraphics[width=\textwidth]{DPP30.png}
\includegraphics[width = \columnwidth]{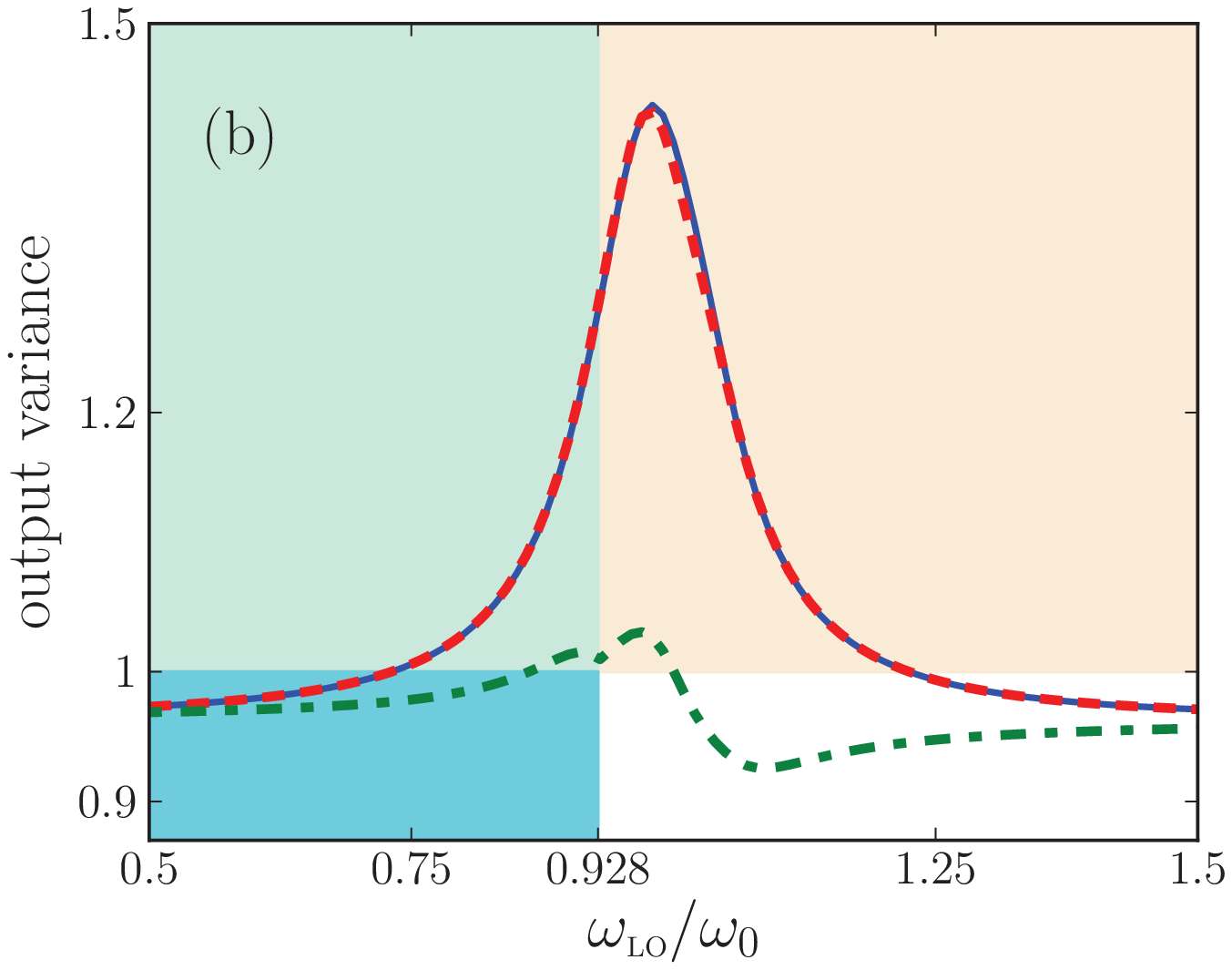}
%\end{minipage}
%
\caption{(Color online) For squeezed light incident at an angle of $\theta=30^\circ$ onto a loss-compensated multilayer metamaterial, a comparison of the predicted variances~(\ref{varianceE}) as would be measured in balanced homodyne detection at a detection angle of also $30^\circ$. The metamaterial and the input states are described in Fig.~\ref{Fig:Scheme of the multilayer loss compensated medium}. Predictions with exact multilayer theory (blue solid line) are compared with the quantum optical effective-index theory (green dash-dotted) and quantum optical effective-medium theory (red dashed), for  $s$-polarized input states of light in panel~(a) and for $p$-polarization in~(b). }
\label{Fig:Comparing the output squeezing 30}
\end{figure*}

We will consider the same metamaterial for which we calculated power spectra before, as detailed in Fig.~\ref{Fig:Scheme of the multilayer loss compensated medium}.
Since the tangential component ${\bf k}$ is preserved under propagation through the multilayer and since there is air on both sides of the metamaterial, the output state of light will emerge from the loss-compensated multilayer at the same angle $\theta$.
Squeezing, specifically quadrature squeezing, occurs when the variance of the quantum fluctuations in one of the quadrature components of the electromagnetic field drop below the vacuum level. Squeezed states have no classical analogues and their non-classicality can be quantified by their associated normally ordered variances of the field operators~\cite{Blow1990a}.
Squeezed light can be produced by transmitting the radiation field through a nonlinear medium with a second-order nonlinearity $\chi^{(2)}$. Mathematically, the squeezed incident quantum states of light can be written as $|L \rangle={\cal S}_\sigma|0 \rangle$ and $|R
\rangle={\cal S}'_\sigma|0 \rangle$, with squeeze operators belonging to a fixed in-plane wavevector ${\bf k}$ given by
\begin{widetext}
\begin{subequations}\label{definition of S for squeezing}
\begin{eqnarray}
{\cal S}_\sigma&=&\exp\left\{\int_0^{\Delta \omega}\mbox{d}\omega\,[\xi^*_\sigma({\bf k},\omega)e^{-{\rm i}\phi_{\sigma,\xi}({\bf k},\omega)}\, a^{(1)\,\dagger}_{\sigma,\,+}({\bf k},\omega)\,a^{(1)\,\dagger}_{\sigma,\,+}({\bf k},2\Omega-\omega)-{\rm h.c.}]\right\},\\
{\cal S}'_\sigma&=&\exp\left\{\int_0^{\Delta \omega}\mbox{d}\omega\,[ \zeta^*_\sigma({\bf k},\omega)e^{-{\rm i}\phi_{\sigma,\zeta}({\bf k},\omega)}\, a^{(N+1)\,\dagger}_{\sigma,\,-}({\bf k},\omega)\,a^{(N+1)\,\dagger}_{\sigma,\,-}({\bf k},2\Omega-\omega)-{\rm h.c.}]\right\}.
\end{eqnarray}
\end{subequations}
\end{widetext}
Here the $a^{(1)}_{\sigma,\,+}({\bf k},\omega)$ and  $a^{(N+1)}_{\sigma,\,-}({\bf k},\omega)$ are
the photonic annihilation operators of the incident fields with polarization $\sigma$ and the transverse wave vector ${\bf k}$ on the left- and right-hand sides of the multilayer slabs, respectively.
It can be seen that the squeeze operators~(\ref{definition of S for squeezing}) correlate pairs of fixed-frequency modes on both sides of the frequency $\Omega$. The amount of squeezing is controlled by the squeeze parameters $\xi_\sigma({\bf k},\omega)$ and $\zeta_\sigma({\bf k},\omega)$, which depend on the frequency,  polarization, and angle of incidence.
We specify the detector to be a balanced homodyne detector. It is well known that squeezing can be measured in such a setup, where the signal field and a strong local oscillator are superimposed on a beam splitter, see Ref.~\cite{Leonhardt:1997a} and the sketch in Fig.~\ref{Fig:Scheme of the multilayer loss compensated medium}. The measured quantity is the difference in the photo currents of two detectors placed in the output arms of the beam splitter, as represented by the operator~\cite{Blow1990a,Artoni1999a}
\begin{equation}\label{operator O}
\hat{O}_\sigma={\rm i}\int_{t_0}^{t_0+T_0}\mbox{d}t\,\{a^{(N+1)\,\dagger}_{\sigma,\,+}
a_{\sigma,\,\rm LO}
-a^\dag_{\sigma,\,\rm LO}
a^{(N+1)}_{\sigma,\,+}
\},
\end{equation}
where on the right-hand side we suppressed the $({\bf k},t)$-dependence of operators for readability. The detector is assumed to be polarization selective, and it is switched on from time $t_0$ to $t_0+T_0$ . The  $a_{\sigma,\,\rm LO}(t)$ and $a^\dag_{\sigma,\,\rm LO}$ are the creation and annihilation
operators of the local-oscillator field with polarization $\sigma$.
This local-oscillator field is assumed to be a single-mode coherent light beam represented by the complex amplitude
$\alpha_{\sigma,\,\rm LO}(t)$ that equals $F_{\rm LO}^{1/2}\exp[{-\rm i}(\omega_{\rm LO}t-\phi_{\sigma,\rm LO})]$,
in terms of a flux $F_{\rm LO}$, a phase $\phi_{\sigma,\rm LO}$, and the frequency  $\omega_{\rm LO}$. With the usual assumption that the local-oscillator field is much more intense than the signal field, the measurement operator $\hat{O}_\sigma$ of Eq.~(\ref{operator O}) can be written as
\begin{equation}\label{definition E for O}
\hat{O}_\sigma=F_{\rm LO}^{1/2}\int_{t_0}^{t_0+T_0}dt\,E_\sigma(\phi_{\sigma,\rm LO},{\bf k},t)
\end{equation}
where the  operator
$E_\sigma(\phi_{\rm LO},{\bf k},t)$ that equals
$a^{(N+1)}_{\sigma,\,+}({\bf k},t)\exp[{\rm i}(\omega_{\rm LO}t-\phi_{\sigma,\rm LO}-\pi/2)] + h.c$
is one quadrature operator of the output field with wave vector ${\bf k}$ and polarization $\sigma$ that exits the  loss-compensated metamaterial on the right in Fig.~\ref{Fig:Scheme of the multilayer loss compensated medium}. Balanced homodyne detection allows to measure a single quadrature component of the scattered field~\cite{Leonhardt:1997a}. From the above definitions,  the variance of the difference photocount in a narrow-bandwidth homodyne detector can be obtained  as~\cite{Artoni1999a,Vasylyev2009a}
\begin{eqnarray}\label{varianceE}
&&\langle[\Delta E_\sigma(\phi_{\sigma,\rm LO},{\bf k},\omega_{\rm LO})]^2\rangle^{\rm out}= \\
&& 1+2\langle a^{(N+1)\,\dagger}_{\sigma,\,+}({\bf k},\omega_{\rm LO}),a^{(N+1)}_{\sigma,\,+}({\bf k},\omega_{\rm LO}) \rangle\nonumber \\
&&+2\,{\rm Re}[\langle a^{(N+1)\,\dagger}_{\sigma,\,+}({\bf k},\omega_{\rm LO}),a^{(N+1)\,\dagger}_{\sigma,\,+}({\bf k},\omega_{\rm LO}) \rangle e^{2{\rm i}\phi_{\sigma,\rm LO}}], \nonumber
\end{eqnarray}
\begin{figure*}[t]
%\begin{minipage}[b]{0.45\linewidth}
%\begin{minipage}[b]{1.0\linewidth}
%\centering
%\includegraphics[width=\textwidth]{DSS60.eps}
%\includegraphics[width=\textwidth]{DSS60.png}
\includegraphics[width = \columnwidth]{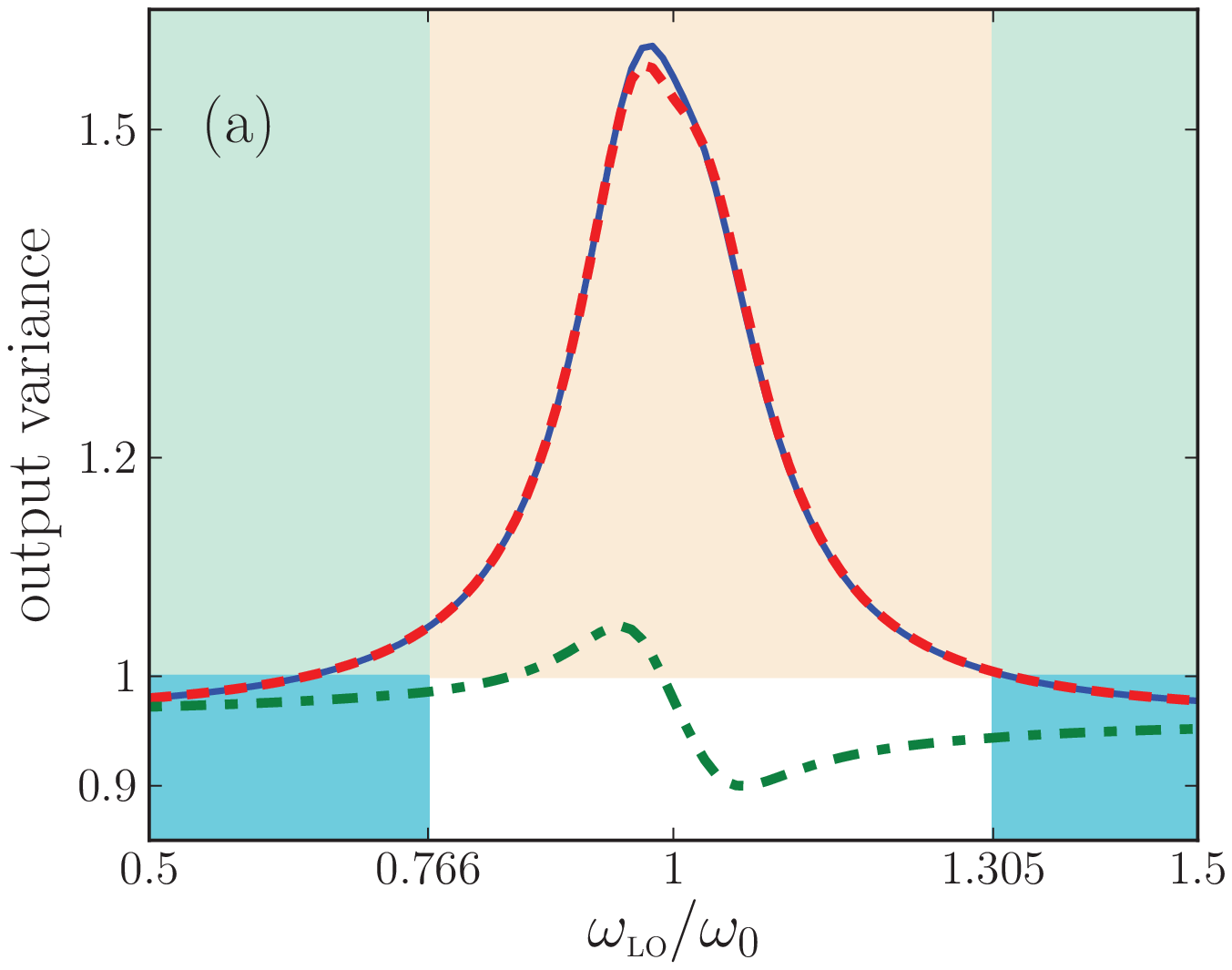}
%\end{minipage}
%\hspace{1.2cm}
%\begin{minipage}[b]{0.45\linewidth}
%\begin{minipage}[b]{1.0\linewidth}
%\centering
%\includegraphics[width=\textwidth]{DPP60.eps}
%\includegraphics[width=\textwidth]{DPP60.png}
\includegraphics[width = \columnwidth]{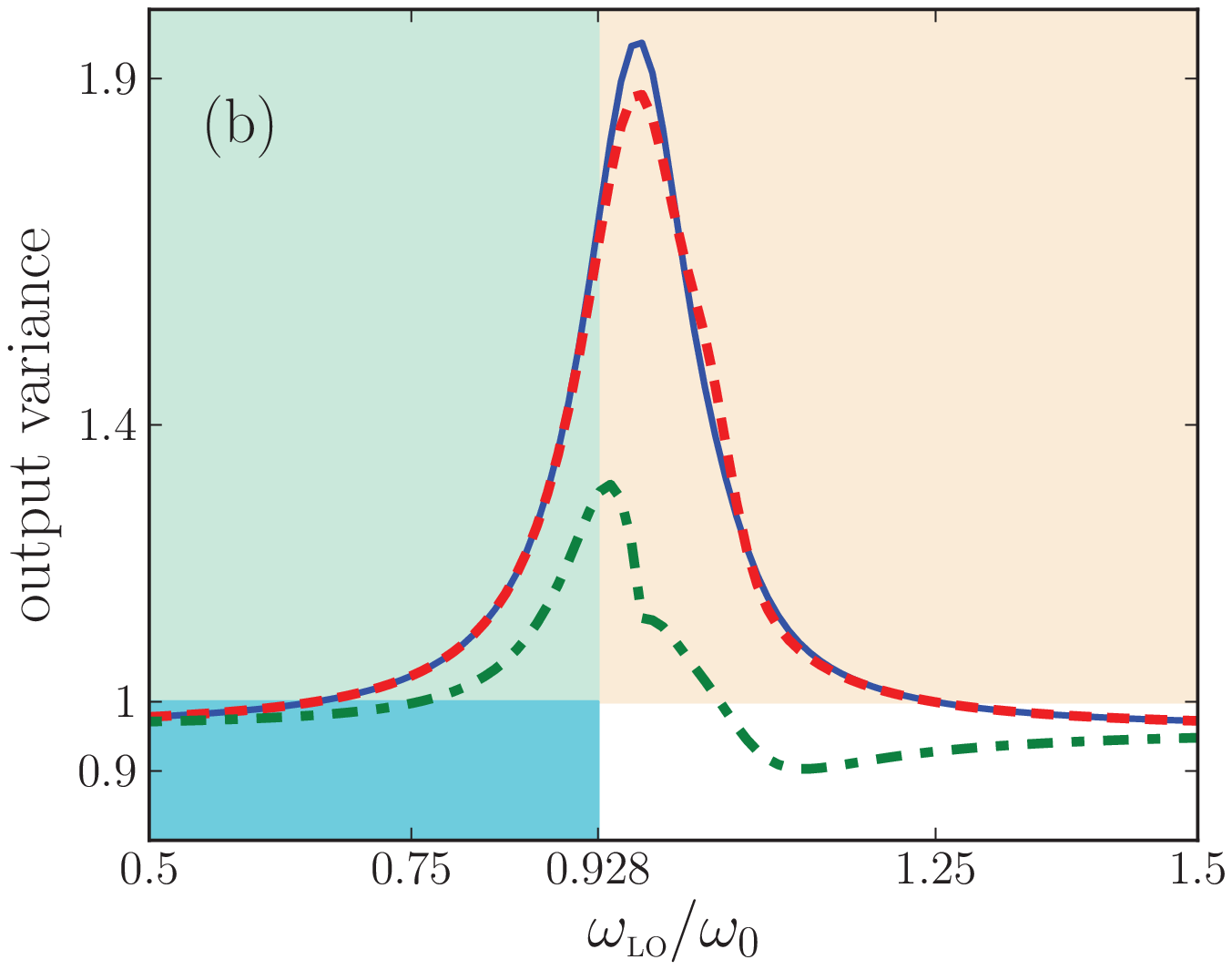}
%\end{minipage}
\caption{Same as Fig.~(\ref{Fig:Comparing the output squeezing 30}) but now for incident and detection angles of  $\theta=60^\circ$.}
\label{Fig:Comparing the output squeezing 60}
\end{figure*}
where the short-hand notation  $\langle C ,D \rangle \equiv \langle C D \rangle -\langle C
\rangle \langle D \rangle$ has been introduced for a correlation. The scattered output state
is squeezed if its photocount variance is smaller than that of the vacuum state value~\cite{Blow1990a}. The homodyne electric-field operator has a variance~(\ref{varianceE}) equal to unity for the vacuum state.  Therefore, the amount of squeezing is gauged by the difference between this variance and unity.
We will now calculate the quadrature variances in Eq.~(\ref{varianceE}) in three ways: using the exact multilayer theory, the effective-index theory, and by the QOEM theory. In all three cases we make use of the commutation relation Eqs.~(\ref{comm_creation_annihilation}) and the definition of the squeezing parameters~(\ref{definition of S for squeezing}). We start calculating the variances~Eq.~(\ref{varianceE})
with the multilayer theory, where the crucial relation between input and output operators is given by ~(\ref{matrix relation for output input}). This will result in rather long expressions, which is one of the reasons to try to find simple but accurate effective theories also in quantum optics.  The two types of correlations in the variance in Eq.~(\ref{varianceE}) are given by
\begin{widetext}
\begin{subequations}\label{eq:expetation_values_for_a_and_a_dagger}
\begin{eqnarray}\label{expectation values for a}
\langle a^{(N+1)\,\dagger}_{\sigma,\,+}({\bf k},\omega),a^{(N+1)}_{\sigma',\,+}({\bf k}',\omega') \rangle &=&\delta_{\sigma \sigma'}\delta({\bf k}-{\bf k}')\delta(\omega-\omega')
\left({|{\cal A}_{\sigma,22}({\bf k},\omega)|^2\, \sinh^2 \xi_\sigma({\bf k},\omega)+ |{\cal A}_{\sigma,21}({\bf k},\omega)|^2\, \sinh^2
\zeta_\sigma ({\bf k},\omega)}\right)\nonumber\\
&&+\langle F_{\sigma,\, +}^\dag({\bf k},\omega)\, F_{\sigma,\,+}({\bf k}',\omega')\rangle,
\end{eqnarray}
\begin{eqnarray}\label{expectation values for a dager}
\langle a^{(N+1)\,\dagger}_{\sigma,\,+}({\bf k},\omega),a^{(N+1)\,\dagger}_{\sigma',\,+}({\bf k}',\omega') \rangle &=& \frac{1}{2}\delta_{\sigma  \sigma'}\delta({\bf k}-{\bf k}')\delta(\omega+\omega'-2\Omega)\\
&&\times\left({\cal A}_{\sigma,22}^{*\,2}({\bf k},\omega)\, \sinh 2\xi_\sigma({\bf k},\omega)e^{-{\rm i}\phi_{\sigma,\xi}({\bf k},\omega)}+ {\cal A}_{\sigma,21}^{*\,2}({\bf k},\omega)\, \sinh 2\zeta_\sigma(\omega)e^{-{\rm i}\phi_{\sigma,\zeta}({\bf k},\omega)}\right).  \nonumber
\end{eqnarray}
\end{subequations}
\end{widetext}

The homodyne signal depends on the noise as described by the operator $F_{\sigma,\,+}({\bf k},\omega)$, which represent the outgoing rightward-propagating noise
field produced inside the multilayer medium. More specifically, the noise dependence is described by the expectation value
$\langle F_{\sigma,\, +}^\dag({\bf k},\omega)\, F_{\sigma,\,+}({\bf k}',\omega')\rangle$, which is the same noise-photon flux that we also came across in the power spectrum~(\ref{Power spectrum of of spontaneous emission}). Thus the effect of the quantum noise on the squeezing properties of output light can be fully characterized by the emitted noise photons. The reason why only the first of these two expressions depends on the quantum noise is that the quantum noise is assumed to be in a thermal state for which $\langle F_{\sigma,\, +}^\dag({\bf k},\omega)\, F_{\sigma,\,+}^\dag({\bf k}',\omega')\rangle$ vanishes.

We will compare predictions of the homodyne signal made with the exact multilayer theory and with the two effective theories.  For the multilayer theory, we can insert into Eq.~(\ref{eq:expetation_values_for_a_and_a_dagger}) the classical multilayer matrix ${\cal A}_{\sigma}({\bf k},\omega)$ of Eq.~(\ref{definition of matrix A}) and the multilayer noise flux  $\langle F_{\sigma,\, +}^\dag({\bf k},\omega)\, F_{\sigma,\,+}({\bf k}',\omega')\rangle_{\rm exact}$ of Eq.~(\ref{correlation for FdagerF}) again.   In both  effective theories on the other hand, the exact matrix coefficients of the input-output matrix are to be  replaced by the corresponding  elements of the effective matrix  $\mathcal{A}_{\rm eff}$ of Eq.~(\ref{eq:Aeffdef}). Furthermore, in the effective-index theory the noise photon flux is given by Eq.~(\ref{eq:F eff}),  while in the QOEM theory it is given by Eq.~(\ref{definition noise for unit cell}).

In Fig.~\ref{Fig:Comparing the output squeezing 30}  we compare the output squeezing spectrum predicted with the three theories, all for $T=0$, at an angle of 30 degrees away from the normal. In Fig.~\ref{Fig:Comparing the output squeezing 60} we show the same for an output angle of 60 degrees.
For simplicity we take the squeezing strengths
$\xi^*_\sigma({\bf k},\omega)$, $\zeta^*_\sigma({\bf k},\omega)$ and phases $\phi_{\sigma,\xi}$, $\phi_{\sigma,\zeta}$ to be constant in the depicted frequency interval.
 We observe that the output squeezing spectrum is sensitive not only to the local-oscillator phase but also to the angle of incidence and the polarization. For this loss-compensated multilayer, the output squeezing spectrum shows a maximum exceeding unity in the vicinity of the resonance frequency. Noise photons destroy the squeezing property of the input field such that the output state will not at all be squeezed for most local-oscillator frequencies in the interval  $[0.5 \omega_{0}, 1.5 \omega_{0}]$ shown in the figures. By contrast, in the same frequency interval the quantum optical effective-index theory predicts the output light to be squeezed for almost all local-oscillator frequencies. In other words, the output state of light of the loss-compensated material is considerably noisier than that of the homogeneous slab with the same $\beta_{\sigma}$.
Thus Figs.~\ref{Fig:Comparing the output squeezing 30} and \ref{Fig:Comparing the output squeezing 60} clearly illustrate the failure of the quantum optical effective-index theory for loss-compensated metamaterials. In Ref.~\onlinecite{Amooghorban2013a} this failure was already established for normal incidence, and here we see that the agreement does not improve when detecting under an angle.
The more important message from the figures  is the very good agreement between the exact theory and QOEM effective theory, not only for normal incidence but also under an angle, and both for $s$- and for $p$-polarized light.
Small numerical differences between the exact theory and the QOEM theory occur only close to resonance and only for large incident angles.

The colors of the frequency intervals in Figs.~\ref{Fig:Comparing the output squeezing 30} and \ref{Fig:Comparing the output squeezing 60} label net loss and net gain, exactly as before in Figs.~\ref{Fig:Comparing the intensity of spontaneous emission 30} and \ref{Fig:Comparing the intensity of spontaneous emission 60}. When loss is exactly compensated by gain, we saw in these earlier figures that  $N_{\rm eff,\, \sigma}({\bf k},\omega,T)$  diverges while the output intensity was continuous. Here in Figs.~\ref{Fig:Comparing the output squeezing 30} and \ref{Fig:Comparing the output squeezing 60} we see that likewise in homodyne detection the output variance is still continuous at those frequencies where $N_{\rm eff,\, \sigma}({\bf k},\omega,T)$  diverges.

\section{Discussion and conclusions}\label{Sec:Conclusions}
We studied the propagation of quantum states of light through metamaterials, and showed that also in quantum optics an effective description of layered metamaterials can be given, for any angle of incidence and polarization. Quantum noise due to material loss or gain has an influence on the  quantum states of light. We showed that in some situations the effective index suffices to describe the quantum noise, while in other cases an additional effective-medium parameter is needed, namely the effective noise-current density.

We tested our quantum optical effective-index theory (one effective parameter) and quantum optical effective-medium theory (two parameters) by calculating spectra and comparing with a full description of the multilayer metamaterial. For loss-compensated metamaterials, the gain regions emit noise photons, not described by the effective-index theory,  that do affect the spectra. They have a similar effect on balanced homodyne detection measurements.  We showed that our quantum optical effective-medium theory describes both the spectra and the homodyne signal well.

For normal incidence we found earlier that quantum noise of passive metamaterials can be described in terms of the effective index, and loss-compensated  require the additional parameter. We now found that this also holds exactly for $s$-polarized light at all angles of incidence, but for $p$-polarized light the additional parameter is also needed for passive systems. For all angles of incidence and polarizations, we derived expressions for the new effective parameter.

Our results can be readily generalized to magnetic layered metamaterials. For metamaterials not composed of multilayers, more work would be needed to derive the effective noise current density. Metasurfaces with gain will similarly require a  description in quantum optics that describes the quantum noise associated with the gain. Another interesting open question is whether the current effective-medium theories suffice to describe higher-order measurements, for example bunching or anti-bunching in intensity correlation measurements, for quantum states of light that propagated through metamaterials.

%%%%%%%%%%%%%%%%%%%%%%%%%%%%%%%%%%%%%%%%%%%%%%%%%%%%%%%%%%%%%%%%%%%%%%%%%%%%%%%%%%%%%%%%%%%%%%%
\section*{Acknowledgments.}
We thank N. Asger Mortensen for stimulating discussions and support.
E.A. wishes to thank the Shahrekord University for support.
M.W. gratefully acknowledges support from the Villum
Foundation via the VKR Centre of Excellence NATEC-II and from the Danish Council for Independent Research (FNU 1323-00087)
The Center for Nanostructured Graphene is sponsored by the Danish National Research Foundation, Project DNRF103.

%%%%%%%%%%%%%%%%%%%%%%%%%%%%%%%%%%%%%%%%%%%%%%%%%%%%%%%%%%%%%%%%%%%%%%%%%%%%%%%%%%%%%%%%%%%%%%%%
\appendix
%%%%%%%%%%%%%%%%%%%%%%%%%%%%%%%%%%%%%%%%%%%%%%%%%%%%%%%%%%%%%%%%%%%%%%%
\section{Methods to obtain classical effective parameters}\label{App:A}
\sssection{Scattering method} The scattering method developed by Smith and
coworkers~\cite{Smith2002a,Smith2006a,Mortensen2010a} has proved
extremely useful. The idea is  to fit the scattering properties of a metamaterial by those of a homogeneous medium, with values for the effective parameters that give the best fit. Finding equivalent bulk parameters in this way is solving an inverse problem. Recently, this approach has been generalized to  obliquely incidence~\cite{Menzel2008a} by assuming that the effective medium can be fully characterized by $\beta_{\rm eff}$, the  wave-vector component in normal direction (here: $\hat z$-direction).  In fact, for oblique incidence, there is no need to introduce the effective refractive
index because all details of wave propagation follow from this parameter $\beta_{\rm eff}$.
What is more, the refractive index may lose its physical meaning and may even become discontinuous, due to the branch cut of the complex square root~\cite{Rockstuhl2008a,Menzel2008a}. By retrieving the normal wave vector component $\beta_{\rm eff}$ from the reflection and transmission coefficients~(\ref{R effsp}) and~(\ref{T effsp}) of a homogenous medium with thickness ${\rm L}$, the index of the plane wave for both polarizations $s$ and $p$ is derived as
\begin{equation}\label{retrived index}
\beta_{\rm eff,\, \sigma}\,{\rm L}=\pm \arccos \left({\frac{1-r^2_{\rm eff,\,\sigma}+t^{'2}_{\rm eff,\,\sigma}
}{2t'_{\rm eff,\,\sigma}}}\right),
\end{equation}
where $t'_{\rm eff,\,\sigma}={\cal A}_{\sigma, 21}\exp({\rm i}\beta_0 {\rm L})$ is the modified
transmission amplitude. In general, the multiple branches associated with the inverse cosine
of Eq.~(\ref{retrived index}) make the unambiguous
determination of the normal wave vector component $\beta_{\rm eff\, \sigma}$ difficult~\cite{Mortensen2010a}.  However, the ambiguity will not arise in our calculations since for simplicity we only consider situations where the wavelength within the medium is much larger than the multilayer length $L$.

\sssection{Dispersion method} Alternatively one can identify values for effective parameters using the  dispersion method:  the effective parameters of a periodic bi-layer system with permittivity functions $\varepsilon_{\rm a}(\omega)$, $\varepsilon_{\rm b}(\omega)$ and thicknesses $d_{\rm a,b}$  are obtained from the Bloch dispersion relation
\begin{eqnarray}
\cos (\beta_{\rm effb,\,s} d)&=&\cos (\beta_{\rm a} d_{\rm a})\cos (\beta_{\rm b} d_{\rm b})\\
&&-\frac{1}{2}\left({\frac{ \beta_{\rm a, \sigma}}{\beta_{\rm b, \sigma} }+\frac{\beta_{\rm b, \sigma} }{\beta_{\rm a, \sigma} }}\right)  \sin (\beta_{\rm a} d_{\rm a})\sin (\beta_{\rm b} d_{\rm b}) \nonumber
\end{eqnarray}
in the long-wavelength limit. We describe $s$- and $p$-polarized light at the same time, since $\beta_{j,\sigma}$ stands for  $\beta_{j,s}=\beta_j$ and $\beta_{j,p}$ for $\frac{\beta_j}{\varepsilon_{j}}$, while $d$ equals the total thickness $d_{\rm a}+d_{\rm b}$ of the two bi-layers. By taking the Taylor expanding them near the point $(\omega,{\bf k})=(0,{\bm 0})$, we obtain the dispersion relation
\begin{equation}\label{Taylor expanding of Bloch index}
\frac{\beta_{\rm effb,\,\sigma}^2}{\varepsilon_{\rm eff,\sigma\, \|}}+\frac{k^2}{\varepsilon_{\rm eff,\sigma\,\bot}}=\frac{\omega^2}{c^2},
\end{equation}
in terms of two important effective parameters, namely $\varepsilon_{\rm eff,s\, \parallel}=\varepsilon_{\rm eff,p\, \parallel}=\varepsilon_{\rm eff,s\, \bot}=(\varepsilon_{\rm a} d_{\rm a} +\varepsilon_{\rm b} d_{\rm b} )/d$ and $\varepsilon_{\rm eff,p\,\bot}=(\varepsilon_{\rm a}  \varepsilon_{\rm b} d)/( \varepsilon_{\rm b} d_{\rm a} +\varepsilon_{\rm a} d_{\rm b})$. These are standard effective dielectric tensor components  that correspond to the direction of the electric field parallel ($\parallel$) and perpendicular ($\perp$) to the layers, respectively.
%
% %%%%%%%%%%%%%%%%%%%%%%%%%%%%%%%%%%%%%%%%%%%%%%%%%%%%%%%%%%%%%%%%%%
\bibliographystyle{wubssty}
\bibliography{EAMWrefs}
%%%%%%%%%%%%%%%%%%%%%%%%%%%%%%%%%%%%%%%%%%%%%%%%%%%%%%%%%%%%%%%%%%%%%%%
%

\end{document}